\documentclass[a4paper,12pt]{article}

\usepackage[utf8]{inputenc}     
\usepackage{graphicx}           
\usepackage{longtable}          
\usepackage{lscape}             
\usepackage{geometry}           
\usepackage{caption}            
\usepackage{booktabs}           
\usepackage{color}              
\usepackage{soul}               
\usepackage{multirow}           
\usepackage{rotating}           

\DeclareUnicodeCharacter{0430}{a}  

\title{A Systematic Literature Review of Unmanned Aerial Vehicles for Healthcare and Emergency Services}
\author{Sara Habibi, Naghmeh Ivaki, João Barata \\
shabibi@dei.uc.pt, naghmeh@dei.uc.pt, barata@dei.uc.pt \\ 
CISUC, Department of Informatics Engineering, University of Coimbra \\ Coimbra, Portugal
}

\begin{document}

\maketitle

\begin{abstract}
    \fontsize{10}{12}\selectfont

Unmanned aerial vehicles (UAVs), initially developed for military applications, are now used in various fields. As UAVs become more common across multiple industries, it is crucial to understand how to adopt them effectively, efficiently, and safely. The utilization of UAVs in healthcare and emergency services has evolved significantly in recent years, with these aerial vehicles potentially contributing to increased survival rates and enhanced healthcare services. 

This paper presents a two-stage systematic literature review, including a tertiary study of 15 review papers and an in-depth assessment of 136 primary publications focused on using UAVs in healthcare and emergency services. The research demonstrates how civilian UAVs have been used in numerous applications, such as healthcare emergencies, medical supply delivery, and disaster management, for diverse use cases such as Automated External Defibrillator (AED) delivery, blood delivery, and search and rescue.

The studies indicate that UAVs significantly improve response times in emergency situations, enhance survival rates by ensuring the timely delivery of critical medical supplies such as AEDs, and prove to be cost-effective alternatives to traditional delivery methods, especially in remote or inaccessible areas. 
The studies also highlight the need for ongoing research and development to address existing challenges, such as regulatory frameworks, security, privacy and safety concerns, infrastructure development, and ethical and social issues. Effectively understanding and tackling these challenges is essential for maximizing the benefits of UAV technology in healthcare and emergency services, ultimately leading to safer, more resilient, and responsive systems that can better serve public health needs.

\vspace{3mm}
    

\vspace{1mm}
    
\end{abstract}

\vspace{3mm}

\setlength{\columnsep}{1cm}

\renewcommand{\tablename}{Table}

\fontsize{10}{12}\selectfont

\section{INTRODUCTION}

Healthcare and emergency systems are crucial for providing timely, life-saving services. Healthcare focuses on medical care, disease prevention, and health promotion, requiring coordination among hospitals, clinics, and community health workers \cite{R005}. Emergency systems respond to critical situations like natural disasters and medical emergencies, operating under limited time frames and challenging conditions. Both demand high efficiency, accuracy, and reliability to ensure quick patient care, integrating various technologies and human resources to manage rapid responses in diverse and high-stress environments \cite{R006}.
UAVs have rapidly emerged as valuable tools across various sectors, offering significant potential for innovation and efficiency. In healthcare and emergency services, UAVs extend beyond traditional boundaries to save lives \cite{R001}. The need for innovative solutions is critical in these sectors due to population aging, longer life expectancies, resource shortages in healthcare facilities, and the urgent need to increase productivity and reduce costs \cite{R003}.

The current healthcare system faces some challenges. Traditional emergency response systems often face delays due to traffic congestion, geographical barriers, and the limited availability of emergency vehicles \cite{R007}. Many rural and remote areas lack sufficient healthcare infrastructure, making it challenging to provide timely medical assistance \cite{R004}. During pandemics, the need for rapid diagnostic preparedness and the safe transportation of medical samples becomes critical \cite{R010}. 

UAVs can address many of these challenges by providing faster emergency response, remote diagnostics, telemedicine, and logistics solutions, such as delivering medicines, defibrillators, organs, blood, and other essential devices\cite{L068, R003}. The use of UAVs for transporting samples can enhance diagnostic preparedness for infectious disease outbreaks, as seen during the Ebola crisis \cite{R012}. Furthermore, UAVs contribute to minimizing human interaction, thus reducing the risk of virus transmission during pandemics \cite{R010}. These advancements have the potential to revolutionize traditional approaches to healthcare delivery, logistics, and telemedicine. As healthcare systems seek to improve accessibility, efficiency, and patient-centered care, the integration of UAVs offers a promising path toward achieving these goals \cite{L007}.

However, despite their great potential, the implementation of UAVs in healthcare services faces several challenges. Technical limitations such as limited battery life and difficulties in transporting heavy loads, regulatory obstacles including licensing and insurance, safety and privacy concerns, particularly in sensitive areas like environmental monitoring and disaster management, the need for reliable and secure infrastructure such as charging stations and communication networks and control systems present significant barriers to the adoption of UAVs \cite{R013}. These challenges have prompted many publications in recent years \cite{R004, L081, L136}, but their results are not yet integrated, and solutions are still lacking.

Additionally, the acceptance of UAV use in emergency and health contexts is influenced by community perceptions and ethical considerations. Studies have shown that public trust and awareness are crucial for the successful implementation of drone technology in these sectors \cite{R011}. 

Addressing these challenges is critical to unlocking UAVs' full potential in transforming healthcare delivery and other vital services.

This study presents a systematic literature review to identify the landscape, gaps, and future opportunities for UAVs in healthcare and emergency services. Our goals are to determine: i) the main application domains and use cases of UAVs in healthcare services, ii) the primary objectives of using UAVs in these sectors, iii) the main challenges and barriers to the adoption of UAVs in these application domains, and iv) the types of UAVs used. A two-stage literature assessment is applied. First, fifteen review papers (also called a tertiary study), published between $2018$ and $2022$, are analyzed to find the gaps and trends in this vibrant field of study.
Secondly, a systematic literature review (SLR) of 136 primary studies was conducted. The sample included papers extracted from the secondary studies (15 review papers) using backward and forward techniques \cite{R003} and a more recent sample of studies addressing UAVs in healthcare for the period $2021-2023$.

This systematic review identifies several key findings regarding the applications, benefits, challenges, and types of UAVs utilized in healthcare services. Firstly, it details the three main areas where UAVs are adopted: delivery of medical supplies, search and rescue operations, and telemedicine. The delivery of medical supplies includes transportation of blood, drugs, lab samples, organs, and vaccines, which can be crucial in both in-hospital and external healthcare settings. UAVs are also vital in search and rescue operations, aiding in disaster monitoring, damage assessment, mapping, situational awareness, and evacuation efforts. Furthermore, UAVs facilitate telemedicine by enabling the transfer of medical imagery and other diagnostic capabilities. The review highlights the advantages of UAVs in healthcare, such as improving response times, increasing survival rates, and enhancing cost-effectiveness. However, it also points out significant challenges, including regulatory obstacles, safety and privacy concerns, and the need for reliable infrastructure and control systems.

The rest of this paper is organized as follows: Section 2 describes the methodology and process followed in this study. Section 3 introduces the result analysis. In Section 4, we discuss the main findings identified during our analysis of the state of the art, in
perspective with the research questions presented earlier. Finally, Section 5 concludes the paper. 

\section{METHODOLOGY}



This section describes the methodology used to perform the systematic literature review on the use of UAVs in emergency services and healthcare. It is based on well-established guidelines for conducting SLRs \cite{R002}, considering two principal stages subsequently explained. 

\subsection{Stage 1: Assessment of Review Studies}

This stage focuses on the identification and assessment of review studies and comprises the following steps: (i) identification of review studies, (ii) study selection, and (iii) data extraction and synthesis of review studies.

\subsubsection{Identification of Review Studies}
\label{sec:Identifications of Studies }
To conduct this step, we determined the main search engines that researchers utilize to locate reliable and peer-reviewed studies that contribute to the specific subject of interest (i.e., using drones in emergency management services and healthcare). Subsequently, we established a general query string to conduct the search.
 
Data selection was done by querying electronic databases such as Google Scholar, PubMed, Scopus, and Web of Science with the following search query string: 

("drone" or "UAV" or "UAS" or “Unmanned aerial vehicle”) and ("emergency" or "health" or "medicine" or "disaster" or "surveillance") and ("review" or "systematic review" or "survey")

The final search, conducted on February 23, 2023, yielded a total of 693 research items.

\begin{figure}
    \centering
     \includegraphics[scale=0.59]{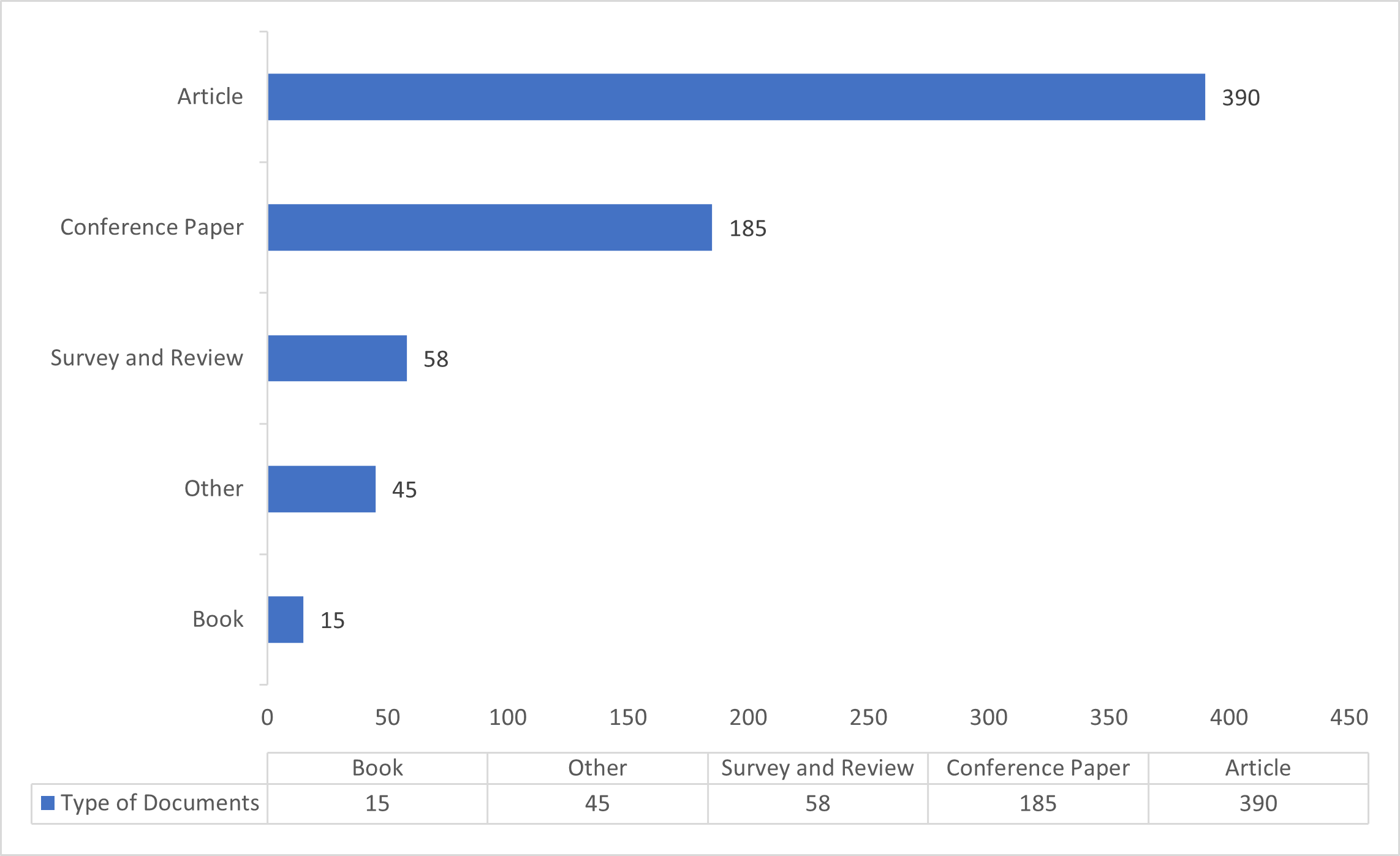}
    \caption[Distribution of Documents by Type]{Distribution of Documents by Type}
    \label{fig:distributionOfReviewPappers}
\end{figure}

\begin{figure}
    \centering
    \includegraphics[scale=0.49]{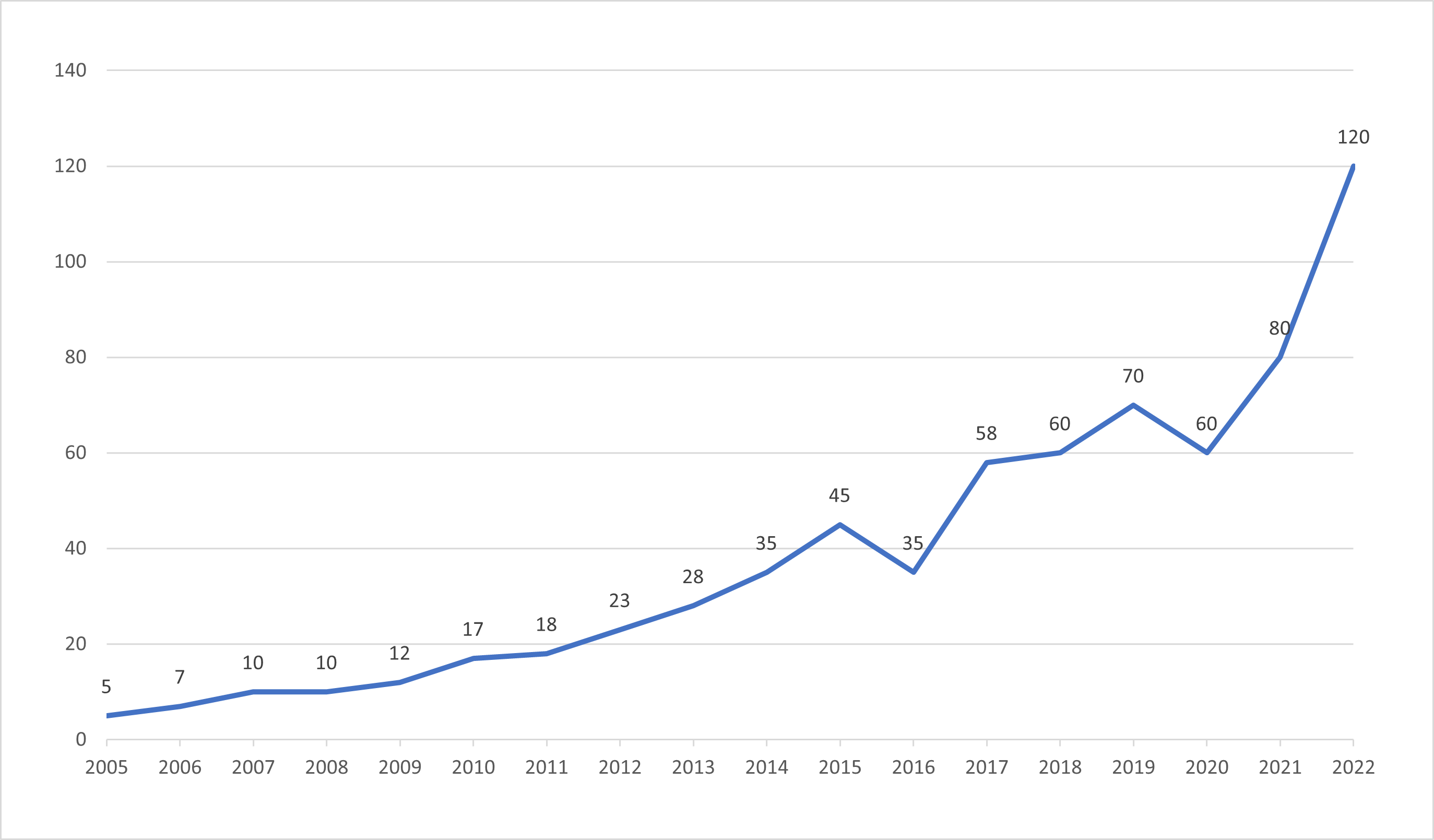}
    \caption[Evolution of Document by Year]{Evolution of Document by Year}
    \label{fig:evilutionOfReviewPappers}
\end{figure}

Figure \ref{fig:distributionOfReviewPappers} and Figure \ref{fig:evilutionOfReviewPappers},  provide an overview of the distribution of the identified research products by document type and year of publication. The breakdown by document type shows a significant number of articles, indicating substantial scholarly work on the subject. The presence of numerous conference papers reflects the nascent characteristics of this research strand. Moreover, the distribution of results over the years shows that publications on this topic were not widespread before 2010. In 2013, NASA developed a conceptual framework for Unmanned Aircraft System (UAS) Traffic Management (UTM) \cite{utm}, which was initially presented at a NASA-industry workshop in 2014. In 2015, NASA hosted a UTM convention where NASA and UAS operators highlighted the need for UAS traffic management. Additionally, in 2016, the first European version of UTM, known as U-Space, and its concept of operations was developed \cite{uspace}. Since then, there has been relatively rapid growth in the number of publications.

\subsubsection{Study Selection}

The 693 identified research products were then evaluated for further analysis. A set of inclusion and exclusion criteria was used to choose the relevant papers. 

\begin{itemize}
    \item Inclusion criteria: 
        \begin{itemize}
            \item Papers published in journals categorized as "survey" or "review".
            \item The study involves the application of UAVs for healthcare purposes.
            \item The study addresses at least one of the following aspects of the utilization of UAVs in healthcare or emergency management services: application of UAVs, challenges, evaluation, or analysis of the regulations governing these devices, emergency management phases, types of UAVs, and adoption of UAVs in healthcare services.
            \item The autonomous vehicles are intended for transporting medical supply deliveries, medicine, and AED deliveries.
        \end{itemize}
    \item Exclusion criteria:
        \begin{itemize}
        
            \item The document language is different from English.
            \item The article was published before 2005.
        \end{itemize}
\end{itemize}

\begin{figure}
    \centering
    \includegraphics[scale=0.40]{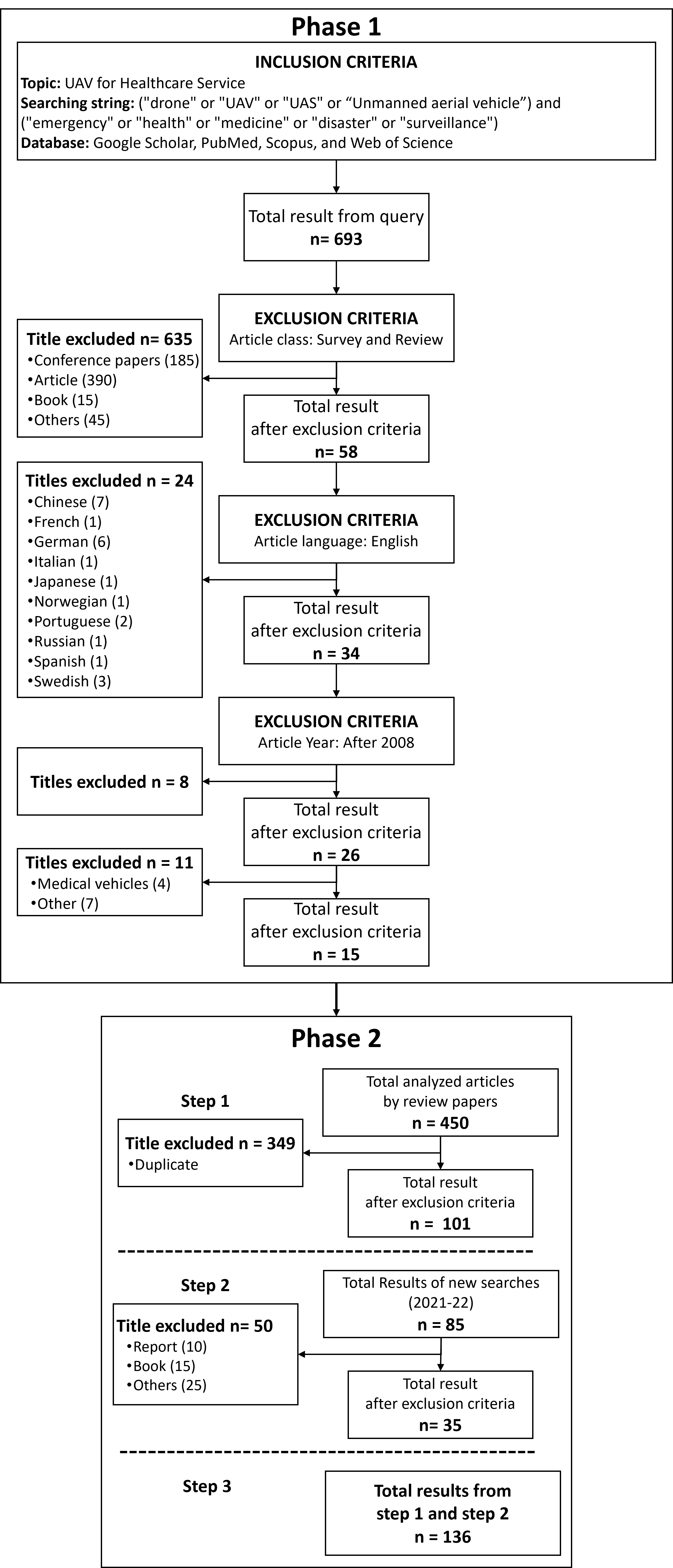}
    \caption[Flowchart mapping the selection process, with the number of identified records included and excluded documents, and the reasons for exclusions.]{Flowchart mapping the selection process, with the number of identified records included and excluded documents, and the reasons for exclusions.}
    \label{fig:researchselection}
\end{figure}

The review papers were assessed based on their titles, abstracts, and full content to determine their quality and adherence to the specified inclusion and exclusion criteria. The initial pool included 390 primary journal papers, 58 reviews, 185 primary conference papers, 15 books, and 45 other types of publications. After applying the criteria, 15 review papers were selected. Figure \ref{fig:researchselection} illustrates the entire selection process and its outcomes.

In the next step, the selected review papers were analyzed to understand the application areas and use cases of UAVs in healthcare and emergency services, as well as the extent to which this technology can provide advantages in these domains. Further analysis of the review papers will be presented in Section \ref{sec:AnalysisofRelatedSurveys}.

\subsection{Stage 2: Assessment of Primary Studies}
\label{sec:assessmentOfPrimaryStudies}

This stage focuses on the identification and assessment of primary studies. It comprises the following steps: (i) definition of research questions, (ii) identification of primary studies, (iii) study selection, and (iv) data extraction and synthesis of selected study papers.

\subsubsection{Definition of Research Questions}

Based on the gaps and limitations identified in the tertiary study phase \cite{R001}, we defined the following research questions that build the objectives of our SLR:

\begin{itemize}
    \item\textbf{RQ1:} What are the main application domains of UAVs in healthcare services and emergency management?
    \item \textbf{RQ2:} What are the main objectives of utilizing UAVs in healthcare services and emergency management?
    \item\textbf{RQ3:} What are the main challenges in adopting UAVs for enhancing healthcare services? 
    \item\textbf{RQ4:} What UAV types are used for healthcare and emergency management services?

\end{itemize}


\subsubsection{Identification of Primary Studies}

To further enhance our understanding of UAV utilization and address the gaps identified in the previous phase, we conducted the second phase of our literature review, focusing on primary studies. Initially, we compiled a dataset of 450 papers analyzed in the review papers previously examined. Out of these, 349 were duplicates, resulting in 101 unique primary papers. Additionally, we performed a search for recent publications (from 2022 to 2023) not covered in the earlier review papers, identifying 84 new documents.

\subsubsection{Study Selection}
\label{sec:Study selection}

After analyzing the new primary studies, we excluded 50 documents that did not meet our selection criteria, resulting in a total of 136 primary papers for further analysis. 
Figure \ref{fig:researchselection} depicts the whole process of study selection and the outcomes.

The selected papers were subsequently analyzed to address the specified research questions. A detailed analysis of these primary papers is presented in Section \ref{sec:AnalysisofPrimaryPapers}.

\section{RESULT ANALYSIS}
\label{sec:Analysis}

This section analyzes the literature, focusing on the selected review papers and primary studies.
\subsection{Analysis of Review Papers}
\label{sec:AnalysisofRelatedSurveys}
Various review studies, including surveys and systematic reviews, address the use of UAVs in healthcare and emergency services domains. Our objective with their assessment was to find answers to our research questions and identify the gaps and limitations.

Table \ref{tab:surveryslist} summarizes the information about the 15 review papers analyzed. The table includes (i) the title, (ii) the reference to the paper, (iii) the year of publication, (iv) the number of primary papers reviewed, (v) the temporal interval of the reviewed papers, and (vi) the main objective of each review. The papers are ordered in the table according to their publication year.

\begin{table}[th!]
\centering
\caption{Review Papers on the Use of UAVs in Healthcare and Emergency Services}
\label{tab:surveryslist}
\resizebox{\linewidth}{!}{%
\begin{tabular}{|p{5cm}|p{1.5cm}|p{2cm}|p{3cm}|p{2cm}|p{2cm}|p{2cm}|p{5cm}|}
\hline
\textbf{Title} & \textbf{Ref.} & \textbf{Journal/ Conference} & \textbf{Journal/ Conference Title}& \textbf{Publication Year} & \textbf{Number of Papers Analyzed}& \textbf{Temporal Scope} & \textbf{Main Objective}\\
\hline
Human drone interaction in delivery of medical supplies: A scoping review of experimental studies	& \cite{R035} & Journal & PLOS ONE & 2022 & 4 & 2016-2020 & Study human-drone interaction \\
\hline
The Role of Drones in Out-of-Hospital Cardiac Arrest: A Scoping Review	& \cite{R036} & Journal & Clinical Medicine & 2022 & 26 & 2016-2022 & Study the role and impact of drones in AED delivery in OHCA \\
\hline
Drone for medical products transportation in maternal healthcare A systematic review and framework for future research	& \cite{R037} & Journal & Medicine & 2020 & 3 & 2009-2019 & Study the adaptation of drones in maternal healthcare \\
\hline
Possibilities of Using UAVs in Pre-Hospital Security for Medical Emergencies & \cite{R038} & Journal & Environmental Research and Public Health & 2022 & 107 & 2017-2022 & Study the applicability of UAVs to support emergency medical systems  \\
\hline
Ethical Considerations Associated with “Humanitarian Drones”: A Scoping Literature Review & \cite{R039} & Journal & Science and Engineering Ethics & 2021 & 47 & 2012-2020 & Study the ethical considerations associated with Humanitarian Drones \\
\hline
Drone technology in maternal healthcare in Malaysia: A narrative review & \cite{R040} & Journal & The Malaysian journal of pathology & 2020 & 25 & 2017-2019 & Study the possible application of drones in improving maternal healthcare in Malaysia \\
\hline
Targeted Applications of Unmanned Aerial Vehicles (Drones) in Telemedicine & \cite{R041} & Journal & Telemedicine journal and e-health & 2018 & 6 & 2008-2017 & Study the application of drones in telemedicine  \\
\hline
Drone Applications for Emergency and Urgent Care: A Systematic Review & \cite{R042} & Journal & Prehospital and disaster medicine & 2022 & 6 & Until 2020 & Study the use of drones in emergency healthcare \\
\hline
Applications of drone in disaster management: A scoping review & \cite{R043} & Journal & Science and Justice & 2022 & 52 & 2009-2020 & Study the applications of drone in disaster management \\
\hline
Current summary of the evidence in drone-based emergency medical services care & \cite{R044} & Journal & Resuscitation plus & 2022 & Not specified & Apr-22 & Study the use of drones in emergency medical services \\
\hline
Remote sensing of natural hazard-related disasters with small drones: Global trends, biases, and research opportunities & \cite{R045} & Journal & Remote sensing of environment & 2021 & 635 & 2006-2021 & Study the use of drones in sensing of natural hazard-related disasters  \\
\hline
The Application of Drones in Healthcare and Health-Related Services in North America: A Scoping Review & \cite{R046} & Journal & MDPI- Drones & 2020 & 29 & 2007-2019 & Study the application of drones in healthcare in North America  \\
\hline
Surgical and Medical Applications of Drones: A Comprehensive Review & \cite{R047} & Journal & JSLS & 2018 & 202 & 2014-2017 & Study the surgical and medical applications of drones \\
\hline
UAV-Based Structural Damage Mapping: A Review & \cite{R048} & Journal & International Journal of Geo-Information & 2019 & Not specified & 2005-2018 & Study the use of drones in structural damage mapping \\
\hline
The use of unmanned aerial vehicles for health purposes: a systematic review of experimental studies & \cite{R049} & Journal & Global health, epidemiology, and genomics & 2018 & 5 & 2008-2018 & Study the use of drones in healthcare  \\
\hline
\end{tabular}
}
\end{table}

Several application domains and use cases are identified in these papers, which are summarized in Table \ref{tab:surveyPapersUseCases}. The table includes (i) a reference to the paper, (ii) the application domain of the papers analyzed in each review (i.e., healthcare emergency or emergency management), (iii) the specific use cases identified, and finally (iv) phases of emergency management covered by the identified use cases. 

\begin{table}[th!]
\centering
\caption{UAVs' Application Domains and Use Cases Covered in Review Papers}
\label{tab:surveyPapersUseCases}
\resizebox{\linewidth}{!}{%
\begin{tabular}{|p{1cm}|p{4.5cm}|p{13.5cm}|p{3cm}|p{3.5cm}|}
\hline
 \textbf{Ref.} & \textbf{Application Domain}& \textbf{Identified Use cases} & \textbf{EMS Phases Covered in the Use Cases}& \textbf{Experimental Environment}\\
\hline
\cite{R035} & Healthcare Emergency & AED Delivery for out-of-hospital cardiac arrest (OHCA) & Response & Simulation \\
\hline
\cite{R036} & Healthcare Emergency & AED Delivery for out-of-hospital cardiac arrest (OHCA) & Response & Mostly Simulation, three interviews \\
\hline
\cite{R037} & Maternal healthcare and obstetric emergencies & Blood Products delivery, Blood Samples Delivery & Response & Simulation \\
\hline
\cite{R038} & Medical Supply Delivery, Healthcare Emergency, Disaster Management, Environmental Monitoring & Vaccines Delivery, AED Delivery for out-of-hospital cardiac arrest (OHCA), Medicines Delivery, Blood products Delivery, Goods Delivery, Search and Rescue, AID Kit Delivery, Detect and Track Fires, Measure the moisture content of soils & Response, Preparedness & Not specified \\
\hline
\cite{R039} & Humanitarian Activities & Not specified & Not specified & Simulation and Real Events \\
\hline
\cite{R040} & Healthcare Emergency & Blood Products Delivery & Response & Not mentioned \\
\hline
\cite{R041} &  Telemedicine, Healthcare Emergency, Medical Supply Delivery & Prehospital Emergency Care, Expediting Laboratory Diagnostic Testing, Vaccines Delivery, AED Delivery for out-of-hospital cardiac arrest (OHCA), Hematological products Delivery, Public health surveillance modality (identification of mosquito habitats as well as drowning victims at beaches), Medical Laboratory Products Delivery & Response, Preparedness & Simulation \\
\hline
\cite{R042} &  Medical Supply Delivery, Healthcare Emergency, Disaster Management & AED Delivery for out-of-hospital cardiac arrest (OHCA), Laboratory samples delivery, Vaccine Delivery, Search and rescue, Triage of people in disaster situations & Response, Preparedness & Simulation \\ \hline

\cite{R043} & Disaster Management & Mapping Damages, Search and Rescue, Transportation, Training & Response & Observational study, Experimental study, Case Study, Interview, Simulation, Questionnaire \\ \hline

\cite{R044} & Healthcare Emergency & AED Delivery for out-of-hospital cardiac arrest (OHCA), Blood products Delivery, Naloxone Delivery, Anti-epileptics Delivery, Epinephrine Delivery & Response & Not specified \\ \hline

\cite{R045} & Disaster Management & Vulnerability assessment and risk modeling, Monitoring barrier, deflection, and retention systems, Hazard detection Preparedness, Provision of baseline data Response, Search and rescue, Evacuation, Temporary clearance of debris, Provision of temporary shelter, Critical infrastructure resumption, Rapid damage assessment, In-depth damage assessment, Infrastructure recovery, Housing recovery, Debris management, Environmental recovery & Mitigation, Preparedness, Response, Recovery & Simulation, real Events \\ \hline

\cite{R046} & Medical Supply Delivery, Healthcare Emergency, Environmental Monitoring, Disaster Management & Medical supplies Delivery (e.g., gauze, testing kits, and medications), AED Delivery for out-of-hospital cardiac arrest (OHCA), Biological samples Delivery (e.g., blood, plasma, organs, and other tissues), Emergency service delivery, Search and rescue, Environmental monitoring (e.g., wildfire, landslide, and air quality monitoring) & Not specified & Not specified \\ \hline

\cite{R047} & Healthcare Emergency, Medical Supply Delivery, Medical surveillance, Telemedicine, Telesurgery, Disaster Management, Environmental Monitoring & Disaster prediction and management, Detection of Harmful substances, Diagnosis and treatment, Telesurgery, Medical transport (Tissue, medication, medical device, patient), AED delivery, Medications Delivery, Blood Products Delivery, Vaccines Delivery, Laboratory Samples Delivery, Pharmaceuticals, Emergency medical equipment delivery, Patient transport, Triage in high-risk environments (Information Gathering about the number of patients in need of care), Detect health hazards (such as heavy metals, aerosols, and radiation), Acquire real-time, high-resolution temporal and spatial information for epidemiology research, Monitoring deforestation, agricultural expansion, and other activities that alter natural habitats and ecological communities, Remote diagnosis and treatment of patients by means of telecommunications technology & Mitigation, Preparedness, Response, Recovery & Not specified \\
\hline

\cite{R048} & Disaster Management & Damage Mapping & Response & Real Events \\ \hline

\cite{R049} & Healthcare Emergency, Disaster Management, Telesurgery & AED Delivery for out-of-hospital cardiac arrest (OHCA), Search and Rescue, Blood Samples Delivery, Helping with surgical procedures in war zones & Response & Experimental pilot study and Simulation \\ \hline

\end{tabular}
}
\end{table}

\begin{table}[th!]
\centering
\caption{Use cases of UAVs and EMS Phases}
\label{tab:usecasesAndEMSPhases} 
\resizebox{\linewidth}{!}{%
\begin{tabular}{|p{5cm}|p{13cm}|p{5cm}|}
\hline
\textbf{Application Domain} & \textbf{Identified Use cases} & \textbf{EMS Phases Covered in the Use Cases} \\
\hline

Healthcare Emergency & AED Delivery for out-of-hospital cardiac arrest (OHCA) & Response \\
 & Biological samples Delivery - Blood, Plasma, or Organ Delivery & Response \\
 & Naloxone Delivery & Response \\
 & Anti-epileptics Delivery & Response \\
 & AID Kit Delivery & Response \\
 & Emergency medical equipment delivery & Response \\
 & Patient Transport & Response \\
 & Epinephrine Delivery & Response \\
\hline
Maternal Healthcare and Obstetric Emergencies & Blood Products delivery & Response \\
 & Blood Samples Delivery & Not Emergency \\
\hline
Medical Supplies Delivery & Laboratory Samples Delivery & Not Emergency \\
 & Gauze and Tissues delivery & Not Emergency \\
 & Testing kits Delivery & Not Emergency \\
 & Medical Devices Delivery & Not Emergency \\
 & Hematological Products Delivery & Not Emergency \\
 & Medications Delivery & Not Emergency \\
 & Medical Laboratory Products Delivery & Not Emergency \\
 & Vaccines Delivery & Not Emergency \\
 & Biological samples Delivery - Organ Delivery & Not Emergency \\
 & Biological samples Delivery - Blood Producst Delivery & Not Emergency \\
 & Biological samples Delivery - Plasma Delivery & Not Emergency \\
\hline
Medical Surveillance & Acquire real-time, high-resolution temporal and spatial information for epidemiology research & Not Emergency \\
 & Detection of harmful substances & Not Emergency \\
 & Detect health hazards (such as heavy metals, aerosols, and radiation) & Not Emergency \\
 & Public health surveillance modality (identification of mosquito habitats as well as drowning victims at beaches) & Not Emergency \\
\hline
Telemedicine & Medical Laboratory Products Delivery for testing & Not Emergency \\
 & Samples delivery for testing and diagnosis & Not Emergency \\
\hline
Telesurgery & Helping with surgical procedures in war zones & Not Emergency \\
\hline
Disaster Management & Disaster prediction and management & Mitigation \& Preparedness \\
 & Monitoring barrier, deflection, and retention systems & Mitigation \& Preparedness \\
 & Hazard detection & Mitigation \& Preparedness \\
 & Provision of baseline data & Mitigation \& Preparedness \\
 & Vulnerability assessment and risk modeling & Mitigation \\
 & Damage Mapping & Response \\
 & Search and rescue & Response \\
 & Evacuation & Response \\
 & Temporary clearance of debris & Response \\
 & Rapid damage assessment & Response \\
 & In-depth damage assessment & Response \\
 & Triage in high-risk environments (Information Gathering about the number of patients in need of care) & Response \\
 & Provision of temporary shelter & Response \& recovery \\
 & Critical infrastructure resumption & Response \& recovery \\
 & Infrastructure recovery & Recovery \\
 & Housing recovery & Recovery \\
 & Debris management & Recovery \\
 & Environmental recovery & Recovery \\
\hline
Environmental Monitoring & Monitoring deforestation, agricultural expansion, and other activities that alter natural habitats and ecological communities & Not Emergency \\
 & Measure the moisture content of soils & Not Emergency \\
 & Detect and Track Fires & Not Emergency \\
 & Environmental monitoring (e.g., wildfire, landslide, and air quality monitoring) & Not Emergency \\
\hline
Humanitarian Activities & not specified & Not Emergency \\
\hline

\end{tabular}
}
\end{table}

It is worth noting that emergency management comprises four phases: (i) Mitigation, which involves efforts to reduce or eliminate the impact of disasters before they occur; (ii) Preparedness, which refers to planning and preparing for potential emergencies; (iii) Response, which refers to immediate actions taken during and after a disaster to ensure safety and reduce damage, and (iv) Recovery, which refers to actions taken to return the situation to normal following a disaster.  Understanding these phases is crucial because they provide a structured approach to handling emergencies effectively, ensuring that all aspects of disaster management are addressed systematically. By recognizing the importance of each phase, stakeholders can better coordinate their efforts, allocate resources efficiently, and improve overall resilience to emergency situations.

We have identified nine distinct application domains, each differing in the type of service provided and the importance of timely delivery. In each application domain, several use cases have been identified, which are summarized in Table \ref{tab:usecasesAndEMSPhases}. 

\textbf{Healthcare Emergency} refers to an urgent situation requiring immediate medical support to prevent serious harm or death. Examples of use cases of UAVs in this application domain are AED Delivery for out-of-hospital cardiac arrest (OHCA) and Blood Products delivery \cite{R035,R036}. 

\textbf{Maternal Healthcare and Obstetric Emergency }refers to the medical care provided to women during pregnancy, childbirth, and postpartum. Obstetric emergencies are critical conditions that arise during pregnancy, labor, or delivery, requiring immediate medical intervention to protect the lives of the mother and baby. One use of the UAVs in this application domain is the rapid delivery of blood or medications during obstetric emergencies \cite{R037}. 

\textbf{Medical Supply Delivery} refers to transporting medical goods, such as medications, equipment, devices, and consumables, from suppliers to healthcare facilities, pharmacies, or directly to patients. Use cases of UAVs in this application domain are laboratory sample delivery, testing kit delivery, medication delivery, plasma, and organ delivery \cite{R041, R042, R046, R047}. 

\textbf{Medical Surveillance} refers to the continuous and systematic collection, analysis, and interpretation of health-related data to plan, implement, and evaluate public health practices. Some use cases for UAVs in this application domain are the detection of health hazards (such as heavy metals, aerosols, and radiation) and the collection of real-time, high-resolution temporal and spatial information at low cost for epidemiology research \cite{R047}. 

\textbf{Telemedicine} refers to using telecommunications technology to provide medical care and consultation at a distance. Examples of use cases of UAVs in this application domain are transporting telemedicine kits to remote patients and enhancing connectivity in remote areas to facilitate virtual consultations \cite{R041}. 

\textbf{Telesurgery} refers to performing surgical procedures at a distance using robotic systems and telecommunications technology. Examples of UAVs' use cases in this application domain are delivering surgical supplies, instruments, and materials and providing communication devices to enable remote surgical assistance \cite{R047}. 

\textbf{Disaster Management} refers to the coordinated efforts to prepare for, respond to, and recover from natural or man-made disasters. Some examples of UAV's use cases in this application domain are damage mapping, search and rescue, and delivering food, water, and medical supplies to affected populations \cite{R043, R048}. 

\textbf{Environmental Monitoring} refers to the systematic collection and analysis of data related to environmental conditions, such as air and water quality, pollution levels, and ecosystem health. Examples of UAV use cases for this application domain are monitoring environmental conditions that impact health, such as air and water quality, and tracking ecosystem changes that could affect public health \cite{R047}. 

\textbf{Humanitarian Activities} refer to the actions taken to provide aid and relief to people in need. Examples include humanitarian aid delivery and ensuring the safety and effectiveness of aid distribution \cite{R039}.

Most of the works covered in these reviews are experimental simulation-based studies \cite{R035, R037, R041, R042, R049}. However, there are also observational studies based on actual events \cite{R043, R039}, interviews \cite{R036, R043}, and questionnaire-based studies \cite{R043}. Emergency use cases are more popular, but the majority address the response phase of EMS \cite{R035, R036, R037, R043, R044, R048, R049}. However, we also found cases that address other phases of EMS \cite{R047, R045}. Table \ref{tab:usecasesAndEMSPhases} presents all use cases covered in the review papers and the corresponding EMS phases in the cases of emergency use cases. 

According to the reviewed papers, the adoption of UAVs in healthcare offers several key benefits. Drones can significantly improve both the quality and accessibility of medical services, particularly for patients facing challenges related to cost, distance, or infrastructure limitations \cite{R041}. By enhancing response times and extending healthcare access to remote or hard-to-reach areas, UAVs have the potential to improve clinical outcomes, such as increasing survival rates after cardiac arrest and severe traumatic injuries \cite{R046}.
However, several challenges were involved, which were highlighted in these review papers. 

\begin{itemize}
    \item Environmental and Climate issues: Geographic features like mountains, bodies of water, snow-covered land, forests, and densely built urban areas relevant issues. Additionally, weather conditions such as wind, rain, and snow affected performance \cite{R046}. Drones typically cannot operate in heavy rain and adverse weather conditions because of the risk of losing communication, decreased aerodynamic efficiency, and reduced operator performance \cite{R037}. 

    \item Safety: Failures in drone operations can damage other vehicles, harm people, and have an environmental impact. Therefore, it is crucial to thoroughly examine the safety of this technology under various conditions \cite{R041}. Additionally, to scale the use of drones, implementing air traffic control and path planning is necessary to ensure safe and efficient operations and to prevent collisions \cite{R038}. Moreover, drone operations should be restricted to critical areas, including airports, government facilities, military facilities, or nuclear sites, to enhance safety and manageability. This implies the requirement for a pre-flight authorization \cite{R041}.
            
    \item Security: Drones are exposed to security attacks and can be used for malicious purposes, such as attacks or spying. In response, initiatives have been undertaken to develop anti-drone guns that can disrupt drones from distances as a safety measure \cite{R041}.

    \item Privacy: Drones can photograph or film people in public places or even their houses, resulting in privacy breaches. Regulations and pre-flight authorization can help to address these concerns \cite{R041}. 

    \item Ethical concerns: Some use cases of drones may have ethical implications, such as humanitarian drones \cite{R039}. Humanitarian drones often collect data, including images and videos, which can raise concerns about the privacy of individuals. Examining the ethical concerns with UAVs is required too \cite{R041}.

    \item Social acceptance: Drones' reputation as weapons and intruders on privacy are relevant barriers to their use beyond the scientific domain and specific pilots \cite{R037}. Moreover, the large-scale use of drones can lead to significant noise pollution \cite{R038}. These factors may impact the social acceptance of this technology. 

    \item Infrastructure: There is a need for a drone hub, a specialized center dedicated to drone-related activities such as launch preparation, maintenance, parcel packaging, and monitoring. However, selecting the right location for such a hub presents significant challenges, particularly in terms of cost, as well as regulatory and legal constraints, such as noise and privacy concerns \cite{R041}.

    \item Limited Power: Due to drones' limited power, it is important to consider several factors when choosing a drone for an operation, such as maximum payload capacity, distance range, and average speed capability \cite{R041}.

    \item Human Skills: Jobs in drone technology encompass various specializations, including technician, operator, external pilot, programmer, and other high-tech functions. Workforces also need skills in management, malfunction detection and resolution, technical equipment use, remote piloting, and mission execution \cite{R041}. Human operator factors, such as pilots’ ability/inability to identify objects through a drone-mounted camera or navigate diverse environments (e.g., forests, snow-covered tundra, or bodies of water), significantly impact drone performance. It is recommended to include specially trained drone pilots in emergency response teams instead of adapting pilot training for existing personnel \cite{R046}.

    \item Costs: Several costs are involved in substituting or complementing existing solutions with drones, including expenses for supporting drone technology, human resources for specially trained pilots, the purchase price of the drones, infrastructure costs (such as landing stations, radar systems, GPS capability, and air traffic control), and the cost of storing and accessing data generated by drones. These costs could become a barrier to the adoption of UAVs in healthcare \cite{R046}.

    \item Regulations: Current regulations and policies governing drone flight operations, payloads, data security, and information privacy are often seen as barriers to the widespread utilization of drones for health interventions. Policymakers must establish formal procedures for the safe use of drones that consider health information privacy and protection \cite{R038,R046}.
    
\end{itemize}

These review papers offer valuable insights, but each has its limitations. Some focus on peripheral topics related to drones, such as human-drone interaction \cite{R035} or ethical considerations \cite{R039}. Others focus on specific drone applications, such as maternal healthcare \cite{R037}, AED delivery \cite{R036}, damage mapping \cite{R048}, or telemedicine \cite{R041}. Some reviews concentrate on particular geographical areas, like North America \cite{R046} or Malaysia \cite{R040}. Additionally, most papers reviewed only a limited number of studies, with exceptions found in papers \cite{R043}, \cite{R038}, \cite{R045}, and \cite{R047}. Finally, other studies focus on the technical aspects of implementing drones in medical security systems \cite{R038}.


\subsection{Analysis of Primary Studies}
\label{sec:AnalysisofPrimaryPapers}

\begin{figure}
    \centering
    \includegraphics[scale=0.6]{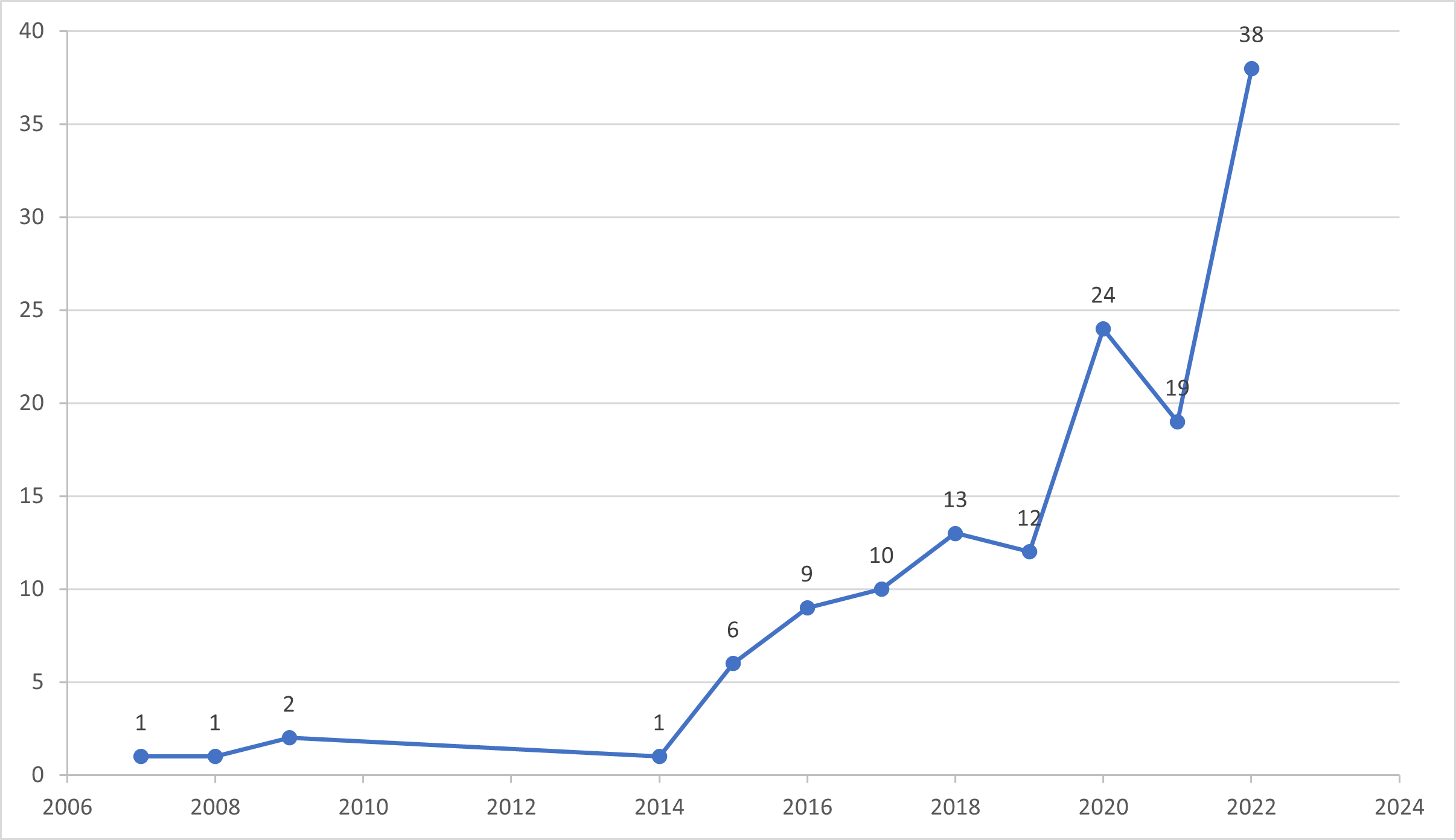}
    \caption[Evolution of analyzed primary research works]{Evolution of analyzed primary research works}
    \label{fig:DistributionofAnalyzedArticlesBasedonTheYear}
\end{figure}

Figure \ref{fig:DistributionofAnalyzedArticlesBasedonTheYear} presents the evolution of papers published in the period 2007 - 2022. The initial years reveal only a few studies published between 2007 and 2014. The next years show exponential growth. A total of 104 of the articles were published within the last five years.

\begin{figure}
    \centering
    \includegraphics[scale=0.46]{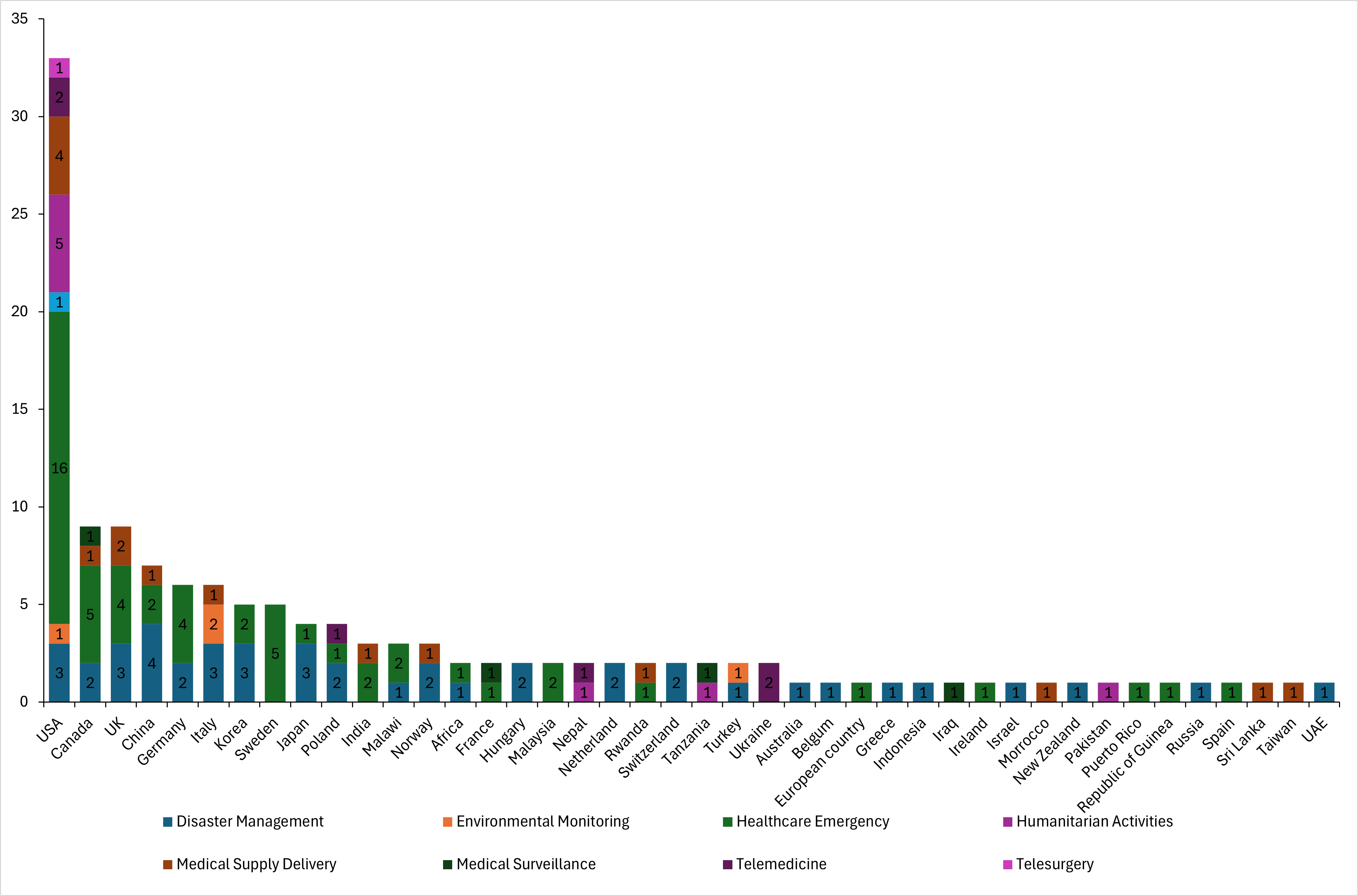}
    \caption[Country and Application Domain]{Country and Application Domain}
    \label{fig:CountryandApplicationDomain}
\end{figure}

Most reviewed papers were conducted in advanced economies like the United States, as illustrated in Figure \ref{fig:CountryandApplicationDomain}, United Kingdom, Canada, China, Germany, and Italy. Sweden and Korea follow in the list with five articles. They are followed by Japan and Poland with four. 
The remaining papers were contributed by India, Malawi, and Norway (3 articles); France,  Hungary, Malaysia, Nepal, the Netherlands, Switzerland, Tanzania, Turkey, Rwanda, and Ukraine (2 articles); and Australia, Belgium, Greece, Indonesia, Iraq, Ireland, Israel, Morocco, New Zealand, Pakistan, Puerto Rico, Republic of Guinea, Romania, Spain, Sri Lanka, Taiwan (1 article).

\subsubsection{Applications Domain of UAVs}

\begin{figure}[th]
    \centering
    \includegraphics[scale=0.60]{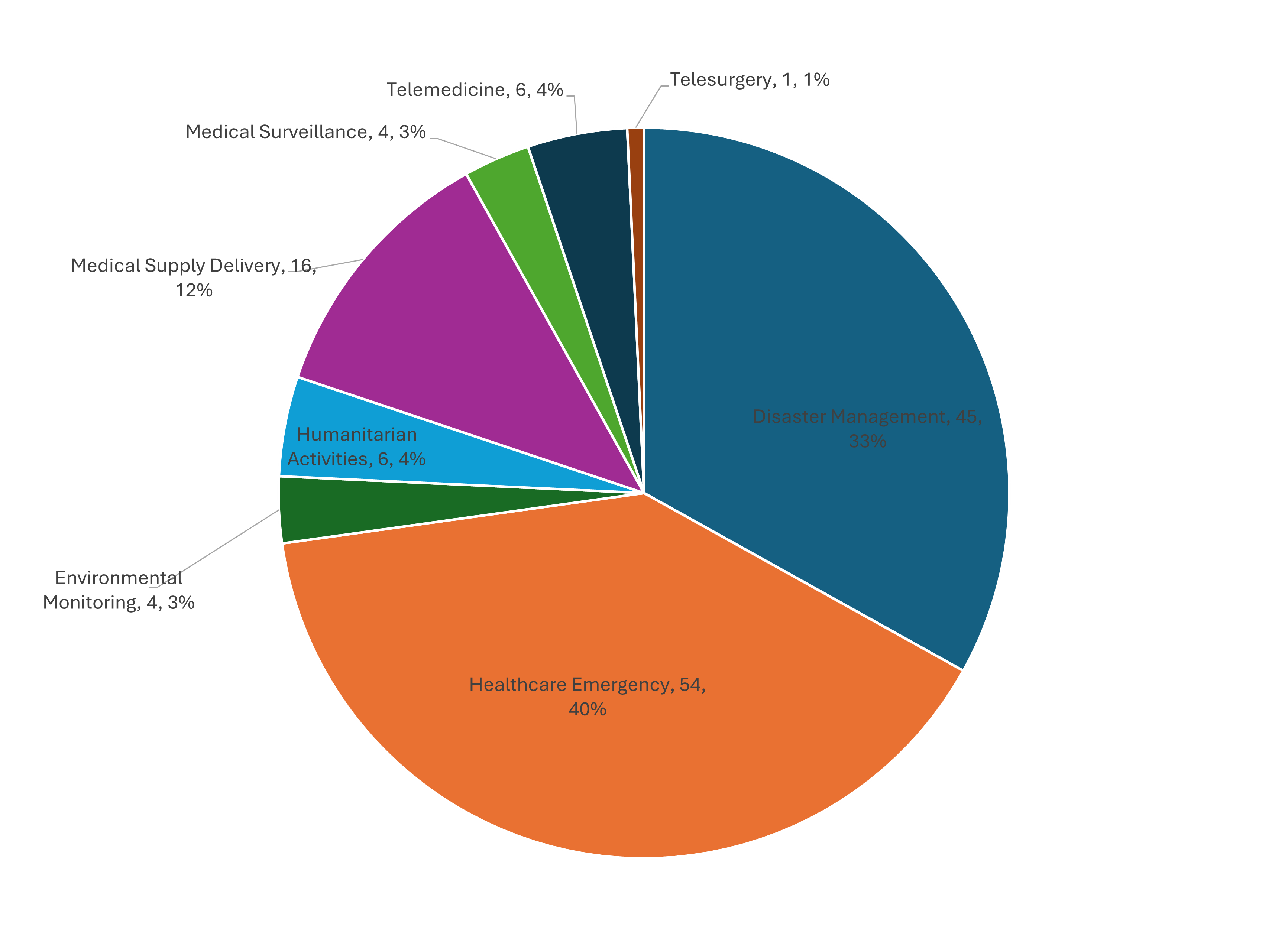}
    \caption[Application Domains of UAVs in Healthcare and Emergency Services]{Application Domains of UAVs in Healthcare and Emergency Services}
    \label{fig:ApplicationDomainwithPercentage}
\end{figure}

Figure \ref{fig:ApplicationDomainwithPercentage} provides an overview of the diverse application domains of UAVs in healthcare and emergency management identified in the analyzed papers, aligning with the application domains identified in the survey papers.

As shown in Figure \ref{fig:ApplicationDomainwithPercentage}, UAVs are predominantly used in healthcare emergencies, representing 40\% of the sample (54 papers). This application domain typically involves rapid AED delivery and emergency medical equipment delivery, such as blood and medications, to patients in urgent need, especially in remote areas. Disaster management is the second most significant application domain, presented in 33\% of the papers (45). UAVs are employed in disaster scenarios for search and rescue operations, damage assessment, risk assessment, and monitoring to enhance the efficiency and speed of disaster response efforts. Medical supply delivery is the third application domain, presented in 12\% of papers (16). In this application domain, UAVs are used to deliver medication, vaccines, and medical equipment. Telemedicine is explored in 6 papers (4\%). This application domain involves using UAVs to support remote medical consultations and the transportation of medical kits, enhancing healthcare access in remote areas. Humanitarian activities, covered by 6 papers (more than 4\%), constitute the next application domain. In this domain, UAVs are utilized to deliver medical aid and supplies to regions affected by conflict or lacking sufficient healthcare infrastructure. In some papers, humanitarian activities are combined with the medical surveillance application domain. Environmental monitoring, with 3\% of papers (4 papers), is the next application domain. In this application domain, UAVs are usually used to track environmental conditions that impact public health, such as air and water quality, and to detect potential health hazards. Similarly, medical surveillance is also presented in 3\% of papers (4 papers). In this application domain, UAVs are used for health monitoring and data collection to manage and prevent disease indications. Finally, Telesurgery, explored in 1 paper (almost 1\%), illustrates the innovative use of UAVs to assist in remote surgical procedures, enabling surgeons to operate from a distance using advanced robotic systems. \\

\noindent \textbf{Healthcare Emergency}\\ \\

The development of UAV technology has revolutionized the speed and efficiency of healthcare responses in emergencies. UAVs are particularly advantageous for the rapid transportation of medical commodities, AED delivery for out-of-hospital cardiac arrest (OHCA), transportation of blood products, emergency medical equipment delivery, and the delivery of critical medications \cite{L147, L128}. For instance, Zipline's drone delivery service, operating in Rwanda and Ghana, reaches over 12 hospitals in Rwanda and operates from six distribution centers in Africa, significantly improving delivery times and the availability of essential medical supplies \cite{L122}. The rapid response capability of Zipline's drones, which can deliver supplies within 15-45 minutes, has been a game-changer in emergency medical services (EMS), where every minute can mean the difference between life and death \cite{L122}. Similarly, the Medication Management System (MMS) in Malaysia employs drones for the swift restocking of medications, ensuring timely delivery, and reducing risks associated with medication delays for elderly patients \cite{L135}. A study in the New England Journal of Medicine highlighted that drone-delivered AEDs can cut delivery time by several minutes, crucial for successful defibrillation and patient survival \cite{L120}.

Beyond speed, UAVs are highly versatile in the types of materials and equipment they can transport. They can bypass traffic and geographical barriers, delivering critical medical supplies swiftly and efficiently \cite{L114}. UAVs directed to specific Ground Control Station (GCS) locations can perform tasks such as acting as access points, relays, or surveillance units, utilizing Software Defined Networking (SDN) for dynamic path and channel selection and Data Distribution Service (DDS) for efficient information dissemination \cite{L155}. These capabilities enable UAVs to maintain robust communication networks in challenging environments. Traditional delivery methods like ground vehicles or helicopters face significant limitations, including traffic congestion and high operational costs. In contrast, UAVs offer a low-cost, efficient alternative that can navigate directly to emergency sites. Research has demonstrated that drones can transport blood products like red blood cells and platelets without compromising their quality, making them an essential option for transporting various medical commodities, including diagnostic samples, medications, vaccines, and blood samples \cite{L158, L154}.

\begin{figure}
    \centering
    \includegraphics[scale=0.60]{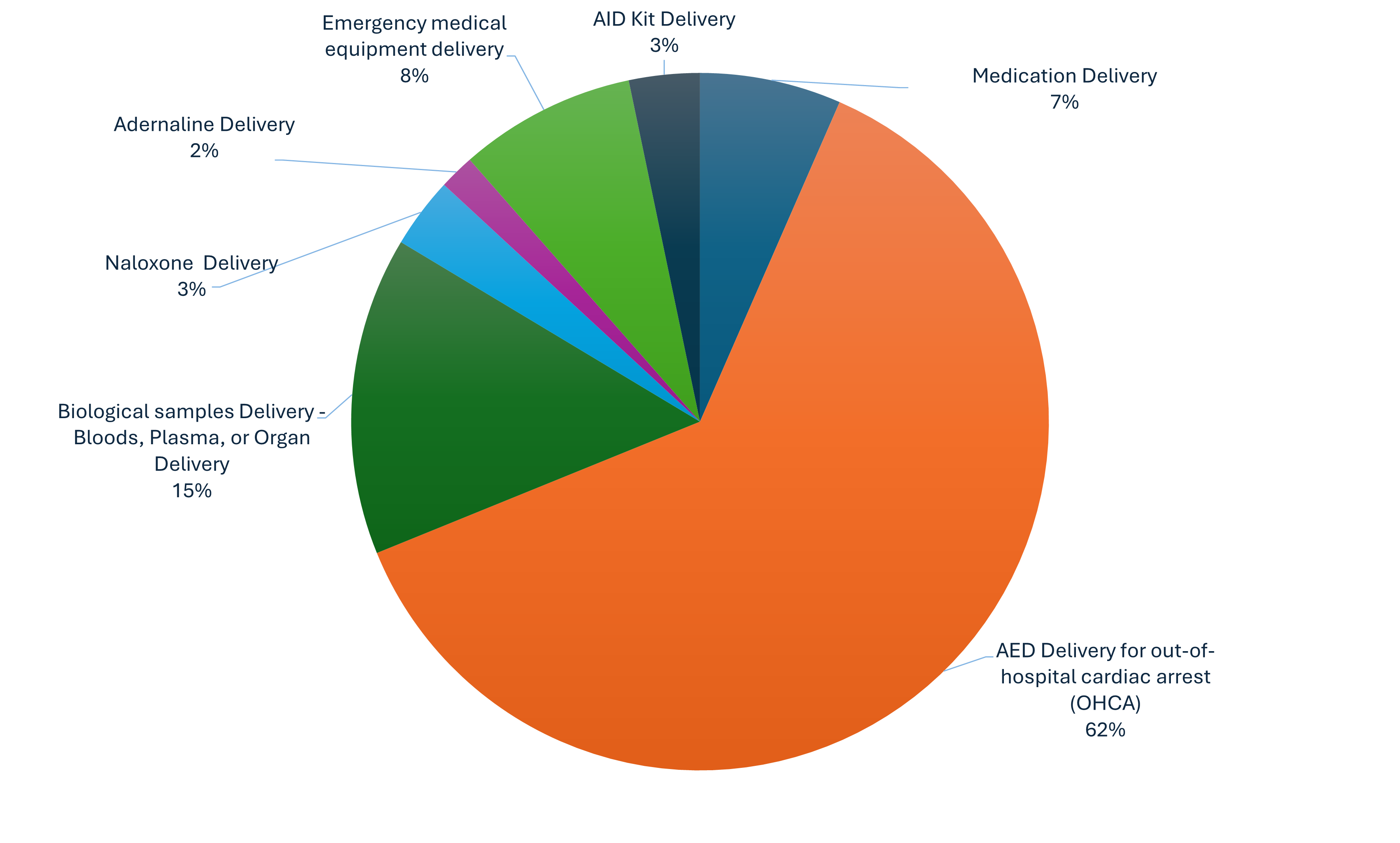}
    \caption[Healthcare Emergency Use cases]{Healthcare Emergency Use cases}
    \label{fig:HealthcareEmergencyUsecases}
\end{figure}

Figure \ref{fig:HealthcareEmergencyUsecases} illustrates the identified use cases in the healthcare emergency domain. The most significant part, accounting for 62\%, is AED delivery for out-of-hospital cardiac arrest (OHCA), emphasizing the crucial need for immediate intervention in cardiac emergencies. 
Biological sample delivery, including blood, plasma, or organ transportation, constitutes 15\% of the use cases and is essential for urgent medical procedures and transplants. Emergency medical equipment delivery, at 8\%, ensures timely access to necessary tools for patient care. Medication delivery represents 7\%, emphasizing the need for quick provision of drugs in urgent situations. Both AID kit delivery and Naloxone delivery account for 3\% each, facilitating prompt first aid and opioid overdose treatment, respectively. Lastly, Adrenaline delivery makes up 2\%, crucial for managing severe allergic reactions. This distribution showcases the vital role of efficient delivery systems in addressing diverse emergency healthcare needs. \\

\noindent \textbf{Disaster Management} \\ \\
UAVs can be used in all stages of disaster management. UAVs can be helpful for disaster prediction by providing real-time data and surveillance capabilities \cite{L085}. For example, monitoring weather patterns, tracking potential hazards, and assessing the risk of natural disasters such as hurricanes, earthquakes, and floods \cite{L093}. UAVs equipped with advanced sensors and imaging technologies can detect hazards such as fires, chemical spills, and radiation leaks \cite{L094}. They can provide baseline data essential for understanding an area's pre-disaster condition, which is critical for effective disaster response and recovery planning \cite{L081}. This data helps in creating detailed maps and models that can be used to assess the extent of damage and prioritize response efforts \cite{L095}. 

UAVs can be utilized for disaster monitoring, providing real-time data to assess the extent of damage and mapping affected areas to facilitate effective response strategies \cite{L089}. They enhance situational awareness by offering aerial views of disaster zones, crucial for planning evacuation routes and delivering aid \cite{L052}. Additionally, UAVs equipped with various sensors and tools, such as cameras and laser equipment, can perform complex tasks during disasters \cite{L143}. The viability and concept of employing a UAV system equipped with a range of sensors, such as a video camera and a tool carrier for equipment like a laser, a release hook, and a searchlight, were also assessed through simulated exercises \cite{L144}. Moreover, with the help of deep learning algorithms, real-time computer vision, and the Global Navigation Satellite System (GNSS), UAVs could soon be able to precisely help first responders \cite{L148}. 

UAVs can also assist in evacuation planning and the temporary clearance of debris \cite{L105}. They can provide real-time aerial views of evacuation routes and identify obstacles that may impede safe evacuation. In high-risk environments, UAVs can be used for triage and gathering information about the number of patients needing care \cite{L167}. They can help identify the most critical cases and ensure that medical resources are deployed where they are most needed \cite{L079}. Additionally, UAVs can assist in providing temporary shelters by identifying suitable locations and delivering essential supplies \cite{L098}. UAVs contribute to housing recovery by assessing the damage to residential areas and identifying safe locations for rebuilding efforts \cite{L166}. They also support environmental recovery by monitoring ecosystems, evaluating the impact of the disaster on natural habitats, and guiding restoration efforts \cite{L164}. UAVs can perform rapid and in-depth damage assessments \cite{L165}. They can be deployed immediately after a disaster to assess structural damages, environmental impacts, and overall devastation \cite{L162}. 

One of the main use cases of UAVs in disaster management involves search and rescue operations. Most studies demonstrated the utility of UAVs in complicated aerial sequences for search, locate, and rescue operations \cite{L100, L099, L101, L164}. It has been demonstrated that using multiple UAVs instead of one can help to effectively search for missing people over a vast region, saving money and time while improving or maintaining the quality of SAR operations \cite{L163}. Multi-object tracking methods were proposed to provide helpful information in complicated serial sequences for victim detection \cite{L060}. These algorithms are based on color and depth data acquired from onboard sensors, integrating photogrammetric and deep learning techniques, placing them using geographic coordinates on DEM and orthomosaic photos, and refining the human identification technique \cite{L091}. This method allowed Search and Rescue (SAR) teams to work more effectively in challenging-to-reach locations.



Researchers have extensively explored the application of UAVs in disaster management. Studies have highlighted various uses, benefits, and challenges associated with UAV deployment in different disaster scenarios. For instance, UAVs have shown potential benefits in hazard monitoring in the Canadian Arctic, although legal and weather restrictions limit their effectiveness in search and rescue or medical response operations \cite{L103}. Additionally, a system utilizing UAVs to locate and communicate with trapped victims via their mobile phones has proven effective in humanitarian disaster scenarios\cite{L164}. The development of a rapid georeference method for UAV images has significantly enhanced post-earthquake response capabilities by enabling the georeferencing of up to one thousand images within a minute\cite{L093}. In emergency scenarios related to rockfall phenomena, the use of micro-UAVs has been advantageous due to their rapid data acquisition, improved safety, and low cost\cite{L090}. UAVs have also been crucial in humanitarian operations, providing valuable sensory data to identify damage and locate missing individuals in events like the Tianjin Port explosion and the Syrian conflict\cite{L031}. Furthermore, UAV photogrammetry has been recognized as a highly effective tool for disaster monitoring and damage assessment, offering a more efficient and accurate alternative to traditional methods \cite{L085}.

\begin{figure}[ht]
    \centering
    \includegraphics[scale=0.60]{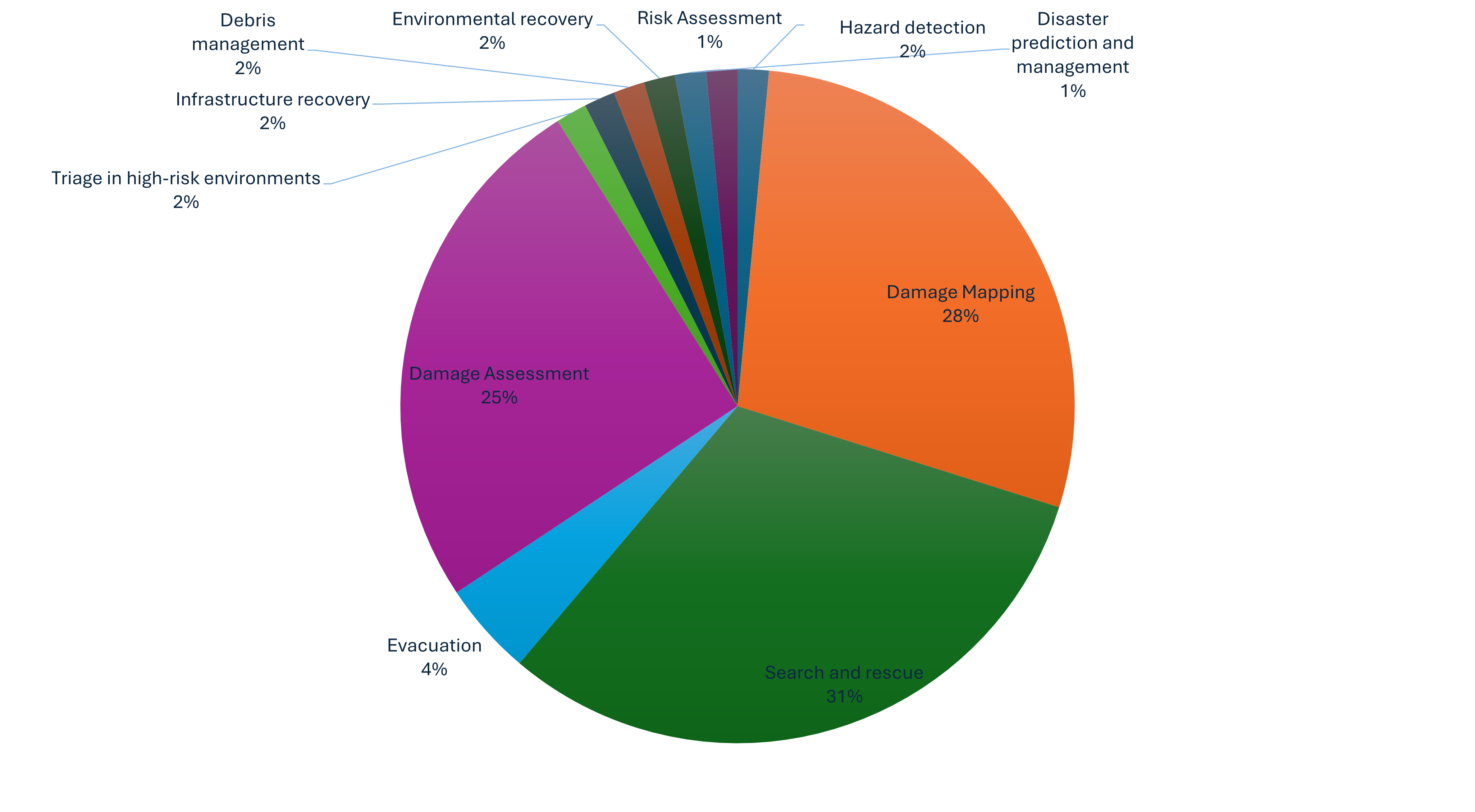}
    \caption[Disaster Management Use cases]{Disaster Management Use cases}
    \label{fig:DisasterManagementUsecases}
\end{figure}

Figure \ref{fig:DisasterManagementUsecases} illustrated the various identified use cases within the disaster management domain. Search and rescue operations account for 31\% of the total use cases, including their critical role in locating and assisting individuals in immediate danger post-disaster. Damage mapping consists of 28\%, highlighting the necessity of assessing damaged areas to allocate resources efficiently. Damage assessment, representing 25\%, focuses on evaluating the severity and impact on infrastructure and the environment, aiding in effective recovery planning. Evacuation procedures, containing 4\%, ensure the safe relocation of people from hazardous areas. Smaller but vital parts include triage in high-risk environments, hazard detection, and environmental recovery, each at 2\%, focusing on prioritizing victims, identifying potential hazards, and restoring affected ecosystems. Risk assessment and disaster prediction and management, both at 1\%, involve evaluating potential risks and planning mitigation and response strategies. Lastly, debris management and infrastructure recovery, each at 2\%, involve clearing debris and rebuilding essential infrastructure. \\

\noindent \textbf{Medical Supply Delivery} \\ \\


UAVs can be used to pick up blood samples from rural homes and deliver prescribed medications based on laboratory results in home health simulations \cite{L138, L150}. For example, Nisingizwe et al. found that the use of UAVs for blood product delivery in Rwanda led to fewer product expirations. Compared with the existing delivery system, the study estimated that products arrived 79–98 minutes earlier than they would have by road. Additionally, the intervention was associated with a 67\% decrease in blood and blood product expiration after the drone delivery program was put in place \cite{L123}.  However, Adrash et al. mentioned several challenges faced by Zipline in implementing their innovative solution in Rwanda, including i)Difficulty in delivery due to the hilly terrain, ii) The limited weight capacity of the UAVs restricts the company to smaller and lighter items, iii) The inability to predict demand, which may result in a lower supply of medical supplies than what is necessary, iv)  Possibility of fraudulent activities and people shooting down the drones, which can increase costs and lead to failure to deliver the supply within the promised timeframe \cite{L122}.

UAVs are becoming popular in low- and middle-income countries (LMICs) that face numerous challenges in getting life-saving vaccines for those needing them \cite{L153}. These challenges include supply chain bottlenecks and inefficiencies, which can cause vaccines to spoil and valuable resources to be wasted before they reach the people who need them \cite{L154}. Additionally, the non-vaccine costs of routine immunization systems are expected to rise by 80\% between 2010 and 2020, with more than one-third of these costs attributable to supply chain logistics \cite{L152}. 

Different studies highlight the potential benefits and challenges of using Unmanned Aerial Systems (UAS) for medical deliveries. Various UAS operational ranges and lifespans were modeled to estimate per-sample costs for lab sample deliveries, determining that the cost-effectiveness of UAS compared to motorcycles is highly sensitive to lifespan and range factors. Longer-range UAS showed potential for cost savings, impacting procurement and maintenance strategies \cite{L073}. Similarly, a computational model assessed the impact of UAVs on vaccine delivery in low- and middle-income countries (LMICs), finding that frequent use of UAVs could enhance vaccine availability and reduce costs by offsetting the initial capital expenses \cite{L074}. The implementation of a highly automated UAV service at San Raffaele Hospital identified significant improvements in hospital logistics efficiency alongside technological and regulatory challenges \cite{L169}. In a different context, the study of UAV transportation of organs demonstrated stable temperature maintenance, minimal vibration and pressure changes, and no damage to the organs, indicating that UAVs could be a reliable alternative to traditional transportation methods, thus improving organ transplantation logistics and outcomes \cite{L129}.

\begin{figure}
    \centering
    \includegraphics[scale=0.60]{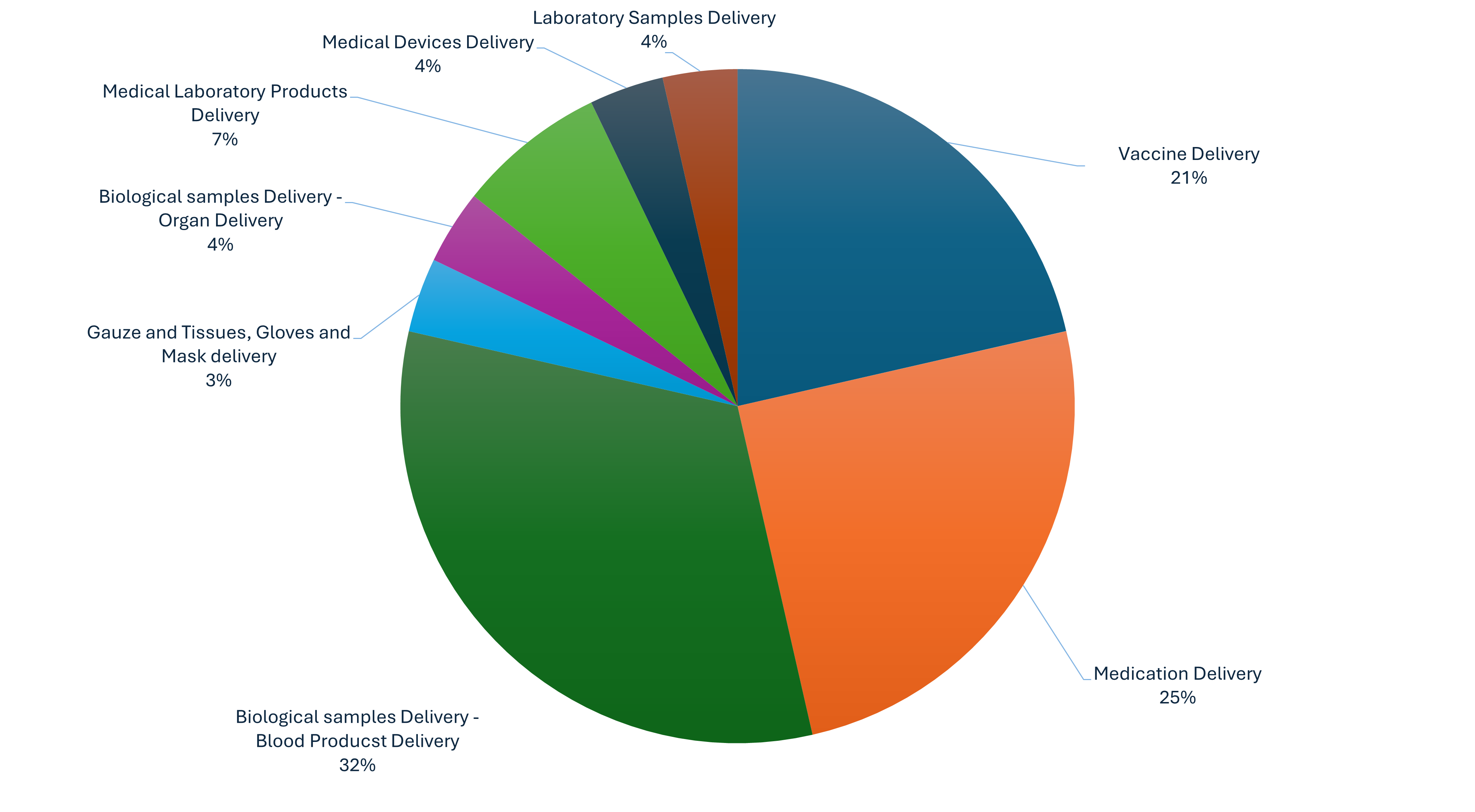}
    \caption[Medical Supply Delivery Use Cases]{Medical Supply Delivery Use Cases}
    \label{fig:MedicalSupplyDeliveryUsecases}
\end{figure}

Figure \ref{fig:MedicalSupplyDeliveryUsecases} presents the identified use cases in medical supply delivery. The most popular use cases (32\%) are dedicated to delivering biological samples, specifically blood products. Vaccine delivery follows at 21\%, emphasizing its significant role in healthcare logistics. Medication delivery is 25\%, highlighting the importance of timely pharmaceutical distribution. Other notable parts include medical laboratory product delivery, which includes 7\% of the use cases. Delivery of medical devices, laboratory samples, and biological samples for organ delivery include 4\% of use cases, and gauze and tissues consist of 3\% of use cases. \\

\noindent \textbf{Telemedicine} \\ \\

Telemedicine integrated with UAVs can open up a new window of remote healthcare and extend its capabilities beyond traditional ways \cite{L065}, but research is still scarce, with only six articles found. One of the essential diagnostic tools that is now accessible through telemedicine and UAVs is cardiac ultrasound. With the ability to transmit real-time cardiac ultrasound images, UAVs enable healthcare providers to remotely assess heart function and detect abnormalities promptly \cite{L136}. The innovative use of UAVs for medical purposes in emergency situations has been discussed, introducing the Healthcare Integrated Rescue Operation (HiRO), which utilizes UAVs to deliver telemedical kits for various medical emergencies such as cardiac events, traumatic injuries, infectious disease outbreaks, and chemical exposures \cite{L067}. 

The potential of UAV-based telemedicine to enhance the safety of first responders and volunteers is highlighted, along with the need for quality and safety assessments to ensure the acceptance and effectiveness of this technology in the emergency management and emergency medical services communities \cite{R043}. Resource-limited settings and pandemic scenarios increase the interest in UAVs for telemedicine, yet it is necessary to incorporate telemedicine competencies into wilderness first responder curricula and fellowships \cite{L056}. Increased attention from the World Health Organization on the development of telemedicine in response to the pandemic is noted, with UAV-supported remote diagnostics and monitoring having the potential to improve patient outcomes and reduce the spread of infectious diseases \cite{L067}. Moreover, integrating UAVs with telemedicine can increase access to specialized medical expertise \cite{L043}. In remote areas or areas with limited access to healthcare professionals or specialists, remote consultations facilitated by UAVs can fill the gap \cite{L056}. Through live video feeds and the transmission of diagnostic images, healthcare providers in hard-to-reach areas can collaborate with specialists located elsewhere to receive expert guidance for complex cases \cite{L171}.

\begin{figure}
    \centering
    \includegraphics[scale=0.60]{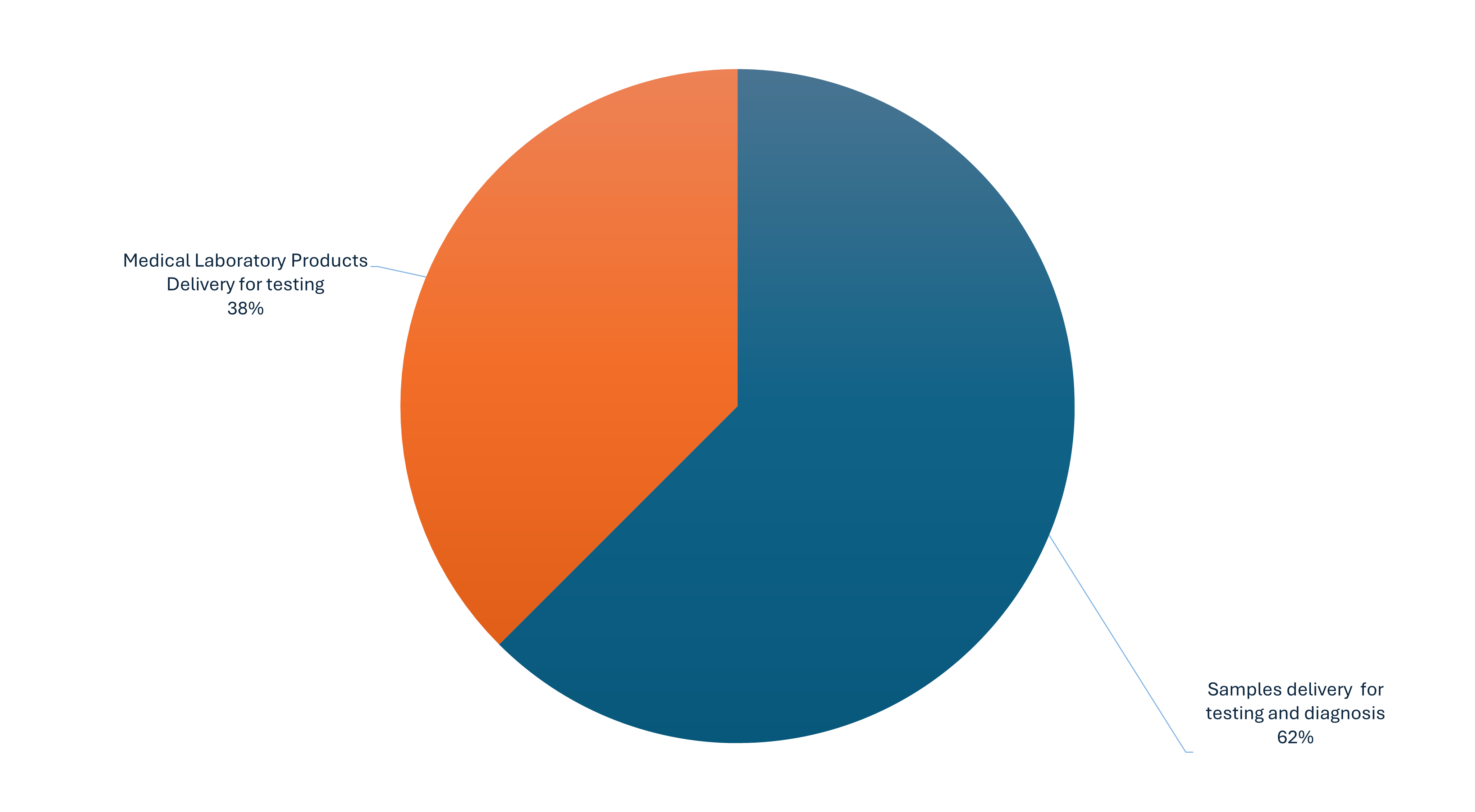}
    \caption[Telemedicine Use Cases]{Telemedicine Use Cases}
    \label{fig:TelemedicineUsescases}
\end{figure}

Figure \ref{fig:TelemedicineUsescases} shows that 62\% of the use cases are for the delivery of samples for testing and diagnosis, indicating a significant emphasis on transporting patient specimens for laboratory analysis. The remaining 38\% of use cases concern the delivery of medical laboratory products for testing, highlighting the critical role of supply logistics in supporting telemedicine services. \\

\noindent \textbf{Humanitarian Activities} \\ \\
Humanitarian activities with UAVs have demonstrated significant potential in enhancing logistics and aid delivery in crises \cite{L059}. UAVs can navigate and deliver aid to areas where traditional transportation methods are prevented by destroyed infrastructure \cite{L064}. For instance, a Continuous Approximation (CA) model was proposed to optimize emergency supply distribution, identifying optimal distribution center locations, service regions, and ordering quantities. The model integrates facility location and inventory management, addressing post-disaster road accessibility and UAV-specific parameters \cite{L048}. 

Another study assessed the perceptions of citizens and government officials regarding UAVs for development activities like mapping flood-prone areas. The study found low awareness of UAV technology among citizens (24\%), but government officials were more familiar with it \cite{L057}. Both groups positively responded to UAVs for disaster relief and infrastructure improvement tasks. Key concerns included security, privacy, and regulation. The study concluded that increased awareness and regulation could help integrate UAVs into humanitarian and development projects in Tanzania \cite{L057}. This trend was coined as "humanitarian neophilia," merging an optimistic belief in technology with market-driven approaches, which was criticized for shifting focus from socio-economic rights and strong state involvement to self-reliance and market solutions \cite{L054}.

\begin{figure}
    \centering
    \includegraphics[scale=0.60]{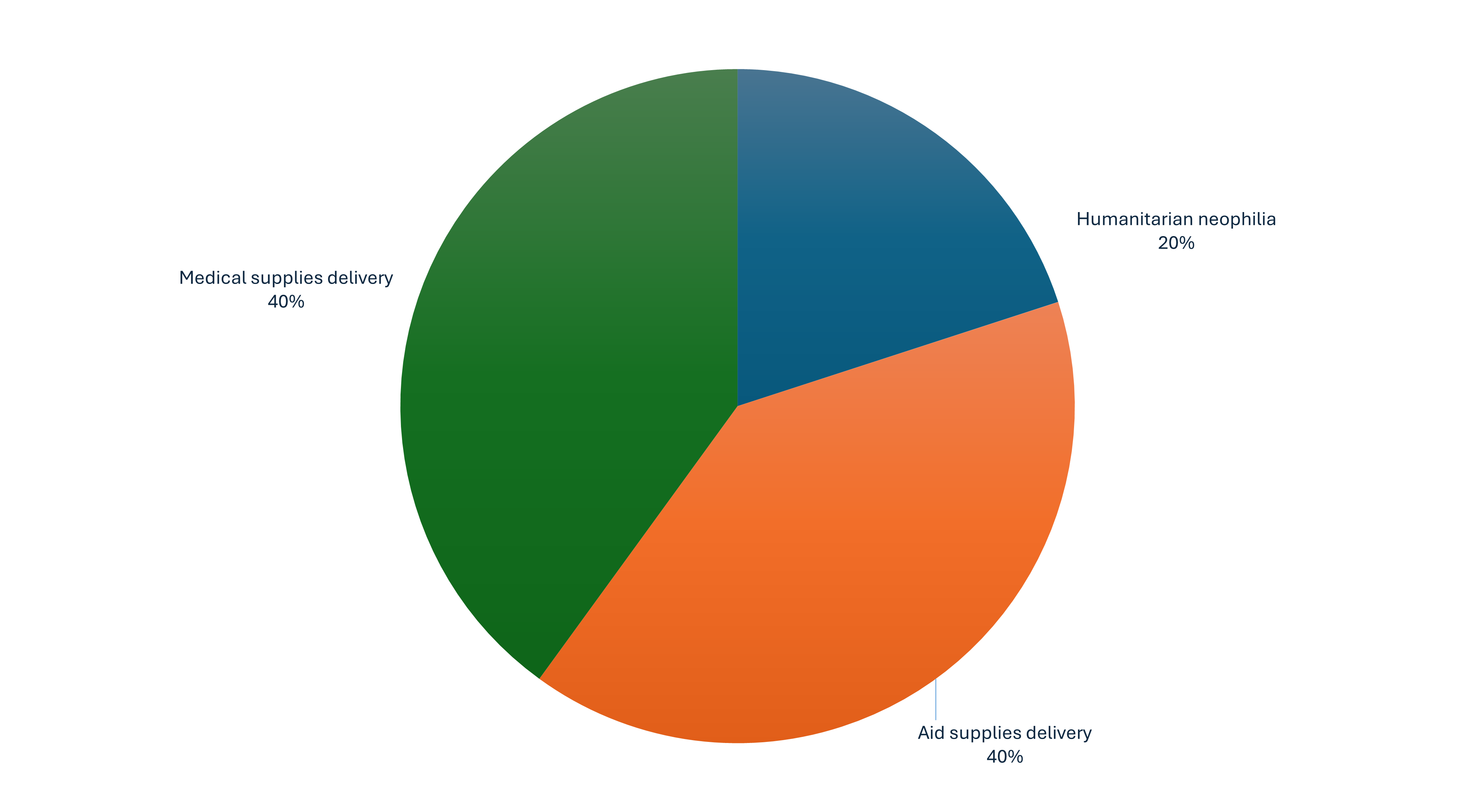}
    \caption[Humanitarian Use Cases]{Humanitarian Activities Use Cases}
    \label{fig:HumanitarianActivitiesUsecases}
\end{figure}

Figure \ref{fig:HumanitarianActivitiesUsecases} illustrates three primary humanitarian activity use cases: medical supplies delivery, aid supplies delivery, and humanitarian neophilia. Each part represents the proportion of the total activities. Medical and aid supply delivery each consists of 40\% of the total activities, indicating a significant focus on delivering essential supplies to those in need. Humanitarian neophilia, which involves interest in new or novel humanitarian methods and technologies, makes up the remaining 20\%. \\

\noindent \textbf{Medical Surveillance} \\ \\
The use of UAVs has become significant in public health monitoring, with notable applications in both the identification of mosquito habitats and aiding visually impaired individuals. UAVs equipped with high-resolution cameras and sensors can survey extensive areas to detect stagnant water bodies, which are prime breeding grounds for mosquitoes. This capability is crucial for controlling mosquito-borne diseases like malaria and dengue, as demonstrated in Zanzibar, where the application of low-cost UAVs, such as the DJI Phantom drone, costing less than $1000$, proved highly effective. These drones captured high-resolution images (7 cm) and generated georeferenced orthomosaics that accurately identified water bodies across various sites, including natural ponds, rice paddies, and urban areas, often surpassing ground surveys in identifying potential mosquito habitats, even those obscured from ground view. This approach provides a flexible, low-cost, and efficient solution for larval source management (LSM) by offering detailed maps of water bodies, access routes, and essential information for planning and implementing vector control measures \cite{L033, L043}. Additionally, UAVs enhance the mobility and safety of visually impaired individuals by detecting safe routes. Equipped with advanced navigation and obstacle detection systems, these UAVs guide individuals along predetermined paths, ensuring their safety and independence \cite{L126}. This dual application of UAV technology underscores its versatility and importance in both disease control and improving the quality of life for individuals with visual impairments.

The use of UAVs to provide remote healthcare services during the COVID-19 pandemic focuses on developing a Portable Health Clinic (PHC) that utilizes a network of self-organizing UAVs to deliver essential healthcare services, such as COVID-19 testing and medication delivery, directly to individuals in hotspot areas under full curfew. This approach minimizes human contact and reduces the risk of virus transmission. The UAVs are equipped with sensors and medications, and they operate autonomously to reach and serve people at their doorsteps. The system follows WHO guidelines and includes a COVID-19 triage process to classify patients based on symptom severity. The PHC system aims to improve the efficiency of healthcare delivery, enhance safety for medical staff, and ensure timely medical assistance during lockdowns \cite{L159}.

\begin{figure}
    \centering
    \includegraphics[scale=0.60]{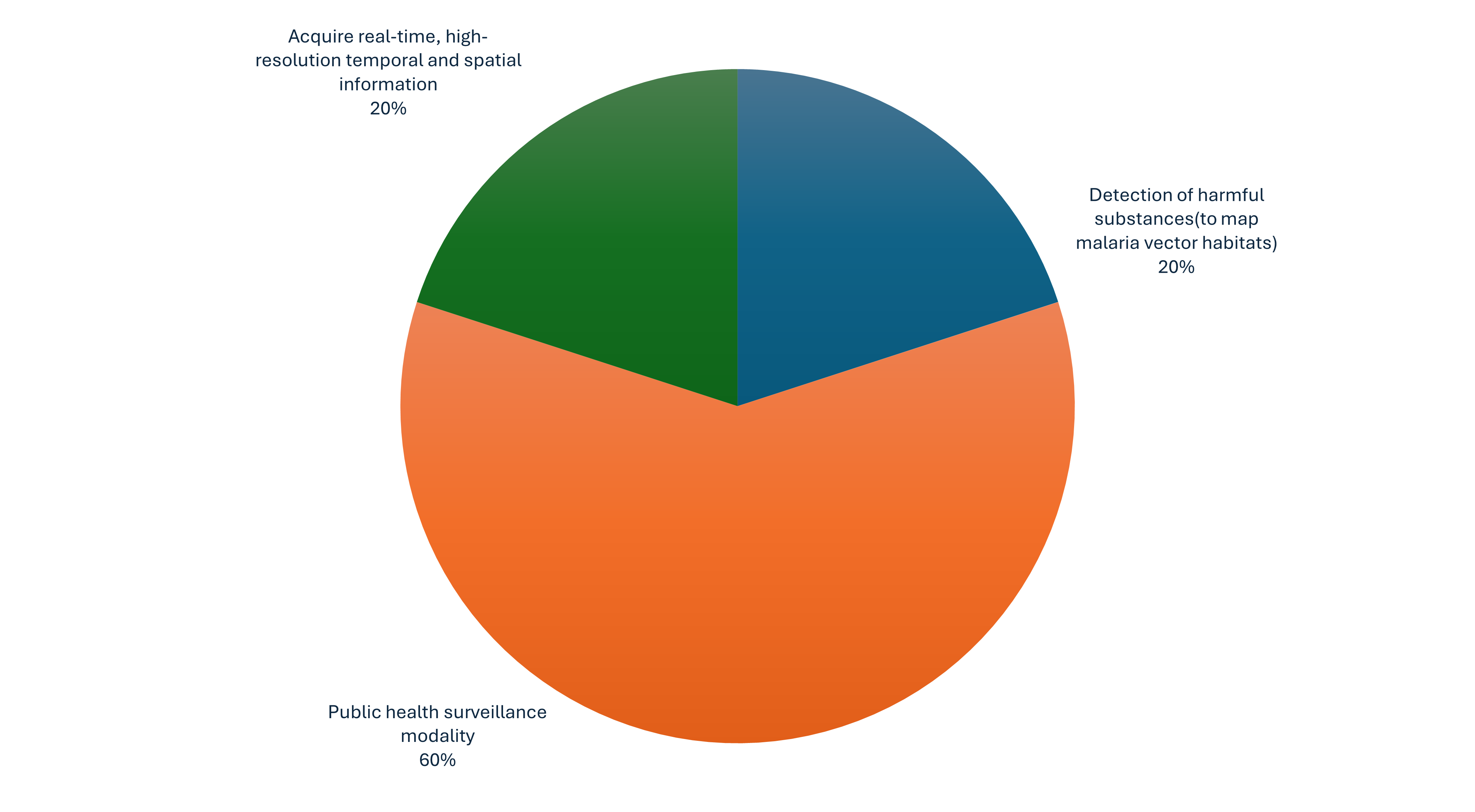}
    \caption[Medical Surveillance Use Cases]{Medical Surveillance Use Cases}
    \label{fig:MedicalSurveillanceUsescases}
\end{figure}

Figure \ref{fig:MedicalSurveillanceUsescases} illustrates the identified use cases in the domain of medical surveillance, highlighting three main categories. The largest part, constituting 60\%, is dedicated to public health surveillance modality, which involves the continuous, systematic collection, analysis, and interpretation of health-related data needed for planning, implementation, and evaluation of public health practice. The second category, making up 20\%, focuses on acquiring real-time, high-resolution temporal and spatial information, crucial for tracking the spread of diseases and responding quickly to health emergencies. The third part, also at 20\%, involves the detection of harmful substances to map malaria vector habitats, which is vital for controlling and preventing the spread of malaria by identifying and managing areas where mosquito vectors grow. \\

\noindent \textbf{Environmental Monitoring} \\ \\
UAVs are crucial in detecting and tracking various ecological hazards, including wildfires, landslides, and air quality issues. In wildfire monitoring, UAVs offer real-time data and high-resolution imagery, which are essential for rapid detection and precise tracking of fire spread, aiding in timely and effective response strategies. This is further enhanced by autonomous tracking methods, where UAVs use distributed control with potential fields to monitor wildfire boundaries. Utilizing a multi-objective evolutionary algorithm, the parameters of these potential fields are optimized to maximize fire coverage while minimizing energy consumption. Simulation results using the FARSITE simulator show that this approach allows UAVs to achieve complete boundary coverage with substantial energy efficiency, highlighting a novel method for distributed autonomy in fire boundary tracking \cite{L034}. For landslides, UAVs can survey unstable terrains and generate detailed topographic maps to predict and monitor landslide activity, significantly improving risk assessment and disaster preparedness \cite{L092}. Additionally, UAVs equipped with air quality sensors can measure pollutants and atmospheric conditions across large areas, providing critical data for air quality monitoring and management. These capabilities underscore the multifaceted utility of UAVs in enhancing ecological hazard detection and management strategies \cite{L035}.. 

\begin{figure}
    \centering
    \includegraphics[scale=0.60]{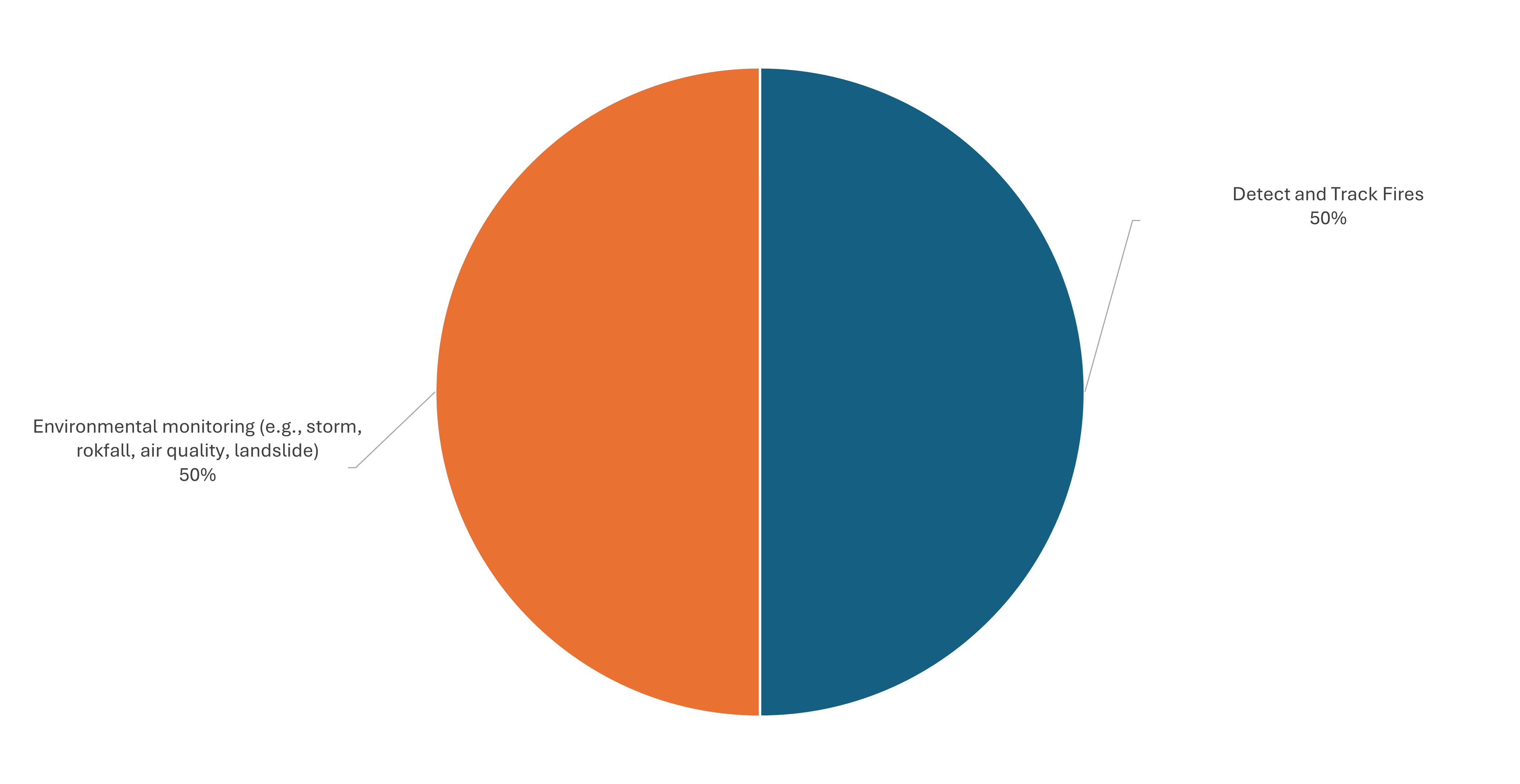}
    \caption[Environmental Monitoring Use Cases]{Environmental Monitoring Use Cases}
    \label{fig:EnvironmentalMonitoringUsecases}
\end{figure}

The application of UAVs in managing rockfall emergencies focuses on the San Germano rockfall event in northwestern Italy, demonstrating how UAVs can be effectively used for rapid and safe data collection in hazardous areas. The UAVs were equipped with high-resolution digital cameras to capture detailed images and create 3D models of the affected area. These models allowed for accurate assessments of the instability and supported emergency response efforts. The research highlights the advantages of using UAVs in emergency scenarios, such as providing comprehensive views of unstable regions and enabling quick decision-making to mitigate risks. The study concludes that UAVs are valuable tools for both qualitative and quantitative evaluations during emergencies, significantly enhancing the management and analysis of geohazards \cite{L090}.
Figure \ref{fig:EnvironmentalMonitoringUsecases} illustrates the identified use cases in environmental monitoring, equally divided between two main categories. The first category, making up 50\%, involves general environmental monitoring for events such as storms, rockfalls, air quality, and landslides. This type of monitoring is crucial for early warning systems and disaster preparedness, helping to protect communities and minimize environmental damage. The second category, also at 50\%, focuses specifically on the detection and tracking of fires. This includes monitoring wildfires and controlled burns, providing real-time data to manage and mitigate their impact, ensuring quick response times, and safeguarding both human lives and natural resources. \\

\noindent \textbf{Telesurgery} \\ \\
 Using telecommunication technologies, UAVs can facilitate real-time, remote surgical interventions, allowing skilled surgeons to operate on patients from thousands of miles away. Exploring the feasibility of using UAVs to facilitate telesurgery in war zones, an experimental mobile robotic telesurgery (MRT) system was demonstrated, utilizing UAVs to establish a network topology capable of supporting surgical operations remotely. Conducted in the challenging environment of the high desert, the project successfully deployed a surgical robot controlled wirelessly via UAVs, enabling remote surgeons to operate on injured soldiers. This setup addresses the logistical and geographical challenges of battlefield operations, providing a robust, low-latency network essential for such critical applications. The study highlights the potential of integrating UAVs into military medical operations, significantly enhancing the capability to deliver urgent and definitive care to soldiers in remote or hostile locations, thereby improving survival rates and reducing risks to medical personnel \cite{L171}.

 \begin{longtable}{|p{4.05cm}|p{8.5cm}|p{3.4cm}|}
    \caption{Application Domain and Use Cases in Primary Papers}
    \label{tab:pp_appDomain} \\
    \hline
    \textbf{Application Domains} & \textbf{Use Cases}&\textbf{Ref} \\
    \hline
    \endfirsthead
    \multicolumn{3}{c}{{\tablename\ \thetable{} -- continued from previous page}} \\
    \hline
    \textbf{Application Domains} & \textbf{Use Cases}&\textbf{Ref} \\
    \hline
    \endhead
    \hline \multicolumn{3}{|r|}{{Continued on next page}} \\ \hline
    \endfoot
    \hline
    \endlastfoot
    
    \multirow{9}{4.05cm}{Disaster Management} & Critical infrastructure resumption & \cite{L166} \\
    \cline{2-3}
    & Damage Assessment & \cite{L080}, \cite{L085}, \cite{L095}, \cite{L133}, \cite{L077}, \cite{L096}, \cite{L093}, \cite{L097}, \cite{L144}, \cite{L032}, \cite{L098}, \cite{L091} \\
    \cline{2-3}
    & Damage Mapping & \cite{L045}, \cite{L078}, \cite{L081}, \cite{L061}, \cite{L089}, \cite{L079}, \cite{L162} \\
    \cline{2-3}
    & Hazard detection & \cite{L094} \\
    \cline{2-3}
    & Human Detection & \cite{L101}, \cite{L099} \\
    \cline{2-3}
    & Disaster prediction and management & \cite{L076} \\
    \cline{2-3}
    & Environmental recovery & \cite{L146}, \cite{L143} \\
    \cline{2-3}
    & Evacuation & \cite{L052}, \cite{L164} \\
    \cline{2-3}
    & Vulnerability assessment and risk modeling & \cite{L060} \\
    \cline{2-3}
    & Search and Rescue & \cite{L100}, \cite{L102}, \cite{L160}, \cite{L103}, \cite{L106}, \cite{L151}, \cite{L163}, \cite{L168}, \cite{L031}, \cite{L040}, \cite{L030}, \cite{L165}, \cite{L105}, \cite{L104}, \cite{L167} \\
    \cline{2-3}
    & Triage in high-risk environments & \cite{L044} \\
    \hline
    
    \multirow{2}{4.05cm}{Environmental Monitoring} & Detect and Track Fires & \cite{L034}, \cite{L035} \\
    \cline{2-3}
    & Environmental monitoring (e.g., wildfire, landslide) & \cite{L090}, \cite{L092} \\
    \hline
    
    \multirow{6}{4.05cm}{Healthcare Emergency} & AED Delivery for out-of-hospital cardiac arrest (OHCA) & \cite{L001}, \cite{L004}, \cite{L005}, \cite{L006}, \cite{L007}, \cite{L008}, \cite{L009}, \cite{L010}, \cite{L011}, \cite{L012}, \cite{L013}, \cite{L014}, \cite{L015}, \cite{L016}, \cite{L017}, \cite{L018}, \cite{L019}, \cite{L020}, \cite{L026}, \cite{L027}, \cite{L072}, \cite{L107}, \cite{L109}, \cite{L110}, \cite{L111}, \cite{L112}, \cite{L113}, \cite{L114}, \cite{L119}, \cite{L120}, \cite{L145}, \cite{L021}, \cite{L022}, \cite{L003} \\
    \cline{2-3}
    & Emergency medical equipment delivery & \cite{L128}, \cite{L149}, \cite{L150}, \cite{L108}, \cite{L140}, \cite{L155} \\
    \cline{2-3}
    & Patient transport & \cite{L137} \\
    \cline{2-3}
    & Adrenaline Delivery & \cite{L118} \\
    \cline{2-3}
    & Biological samples Delivery, Bloods or Organ Delivery & \cite{L028}, \cite{L122}, \cite{L129}, \cite{L050}, \cite{L051}, \cite{L073}, \cite{L161}, \cite{L065}, \cite{L063}, \cite{L121} \\
    \cline{2-3}
    & AID Kit Delivery & \cite{L135} \\
    \cline{2-3}
    & Naloxone Delivery & \cite{L115}, \cite{L116} \\
    \hline
    
    \multirow{2}{4.05cm}{Humanitarian Activities} & Aid supplies delivery, Medical supplies delivery & \cite{L048}, \cite{L059} \\
    \cline{2-3}
    & Humanitarian neophilia & \cite{L054}, \cite{L057}, \cite{L064} \\
    \hline
    
    \multirow{6}{4.05cm}{Medical Supply Delivery} & Biological samples Delivery, Blood Products Delivery & \cite{L123}, \cite{L124}, \cite{L158}, \cite{L125}, \cite{L170} \\
    \cline{2-3}
    & Biological samples Delivery, Organ Delivery & \cite{L148} \\
    \cline{2-3}
    & Medication Delivery & \cite{L068}, \cite{L147}, \cite{L138} \\
    \cline{2-3}
    & Medical Laboratory Products Delivery & \cite{L069}, \cite{L169} \\
    \cline{2-3}
    & Vaccine Delivery & \cite{L074}, \cite{L152}, \cite{L156}, \cite{L154} \\
    \cline{2-3}
    & Gauze and Tissues, Gloves and Mask delivery & \cite{L153} \\
    \hline
    
    \multirow{2}{4.05cm}{Medical Surveillance} & Detection of harmful substances & \cite{L049} \\
    \cline{2-3}
    & Public health surveillance modality & \cite{L033}, \cite{L126}, \cite{L159} \\
    \hline
    
    \multirow{1}{4.05cm}{Telemedicine} & Medical Laboratory Products Delivery for testing & \cite{L043}, \cite{L046}, \cite{L056}, \cite{L067}, \cite{L136}, \cite{L157} \\
    \hline
    
    \multirow{1}{4.05cm}{Telesurgery} & Helping with surgical procedures in war zones & \cite{L171} \\
    \hline
    
    \end{longtable}

\subsubsection{UAVs and Emergency Phases}

Emergency management comprises four phases: the pre-disaster phases, which include \textbf{mitigation} (i.e., preventing emergencies or minimizing their effects) and \textbf{preparedness} (i.e., preparing to handle an emergency); the disaster phase, which involves \textbf{response} (i.e., quick and safe reaction to an emergency); and the post-disaster phase, which consists of \textbf{recovery} (i.e., recovering from an emergency). Each phase presents unique challenges and requires appropriate management strategies. 

\begin{figure}
    \centering
    \includegraphics[scale=0.6]{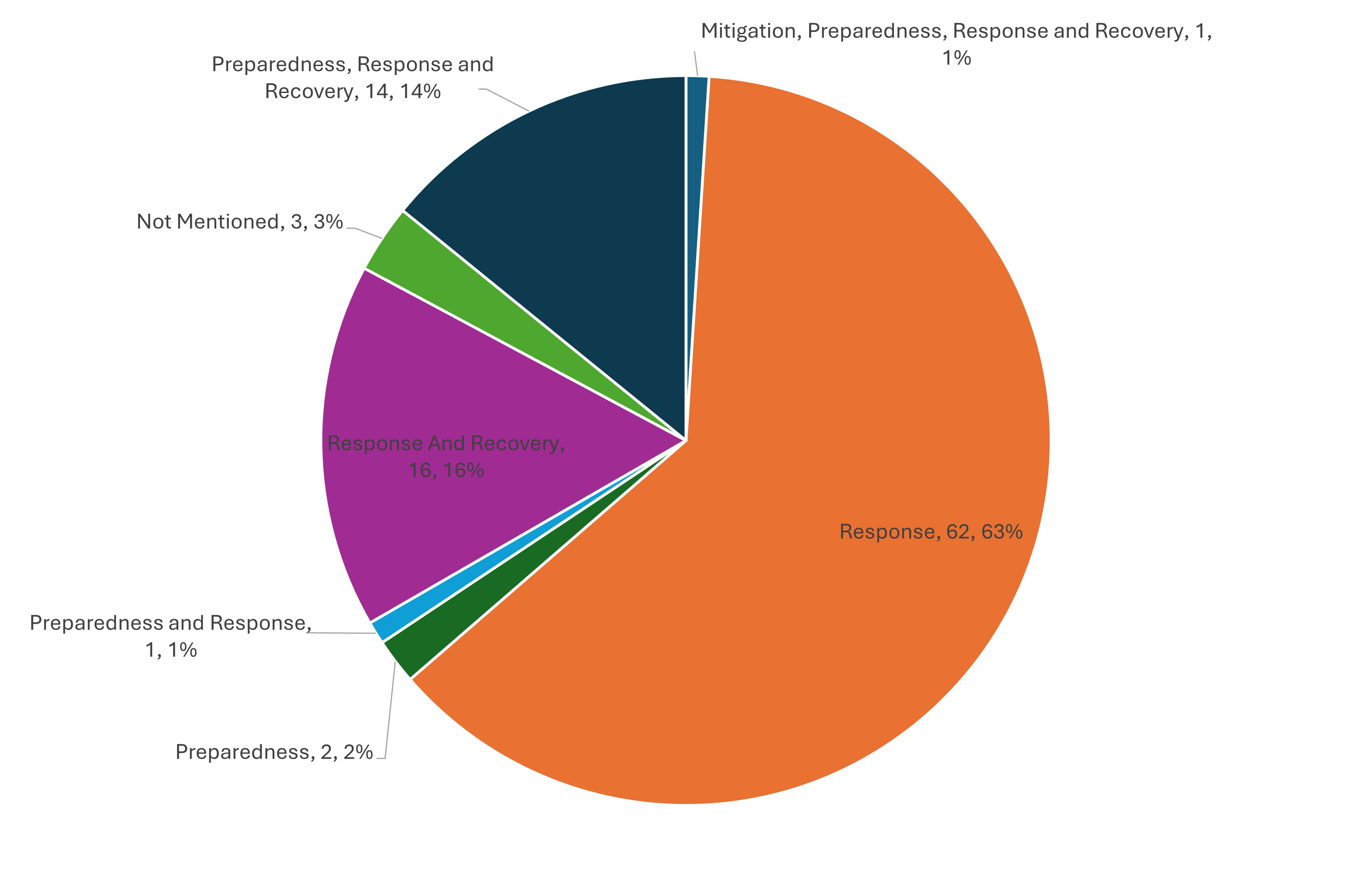}
    \caption[Distribution of Emergency Phases]{Distribution of Emergency Phases}
    \label{fig:EmergencyPhases}
\end{figure}

Figure \ref{fig:EmergencyPhases}, illustrates the distribution of 99 out of 136 studies that focused on disaster management and healthcare emergency use cases. The majority of 62 studies (63\%) focused on the response phase. The second-highest contribution was in the both Response and Recovery, with 16 articles (16\%). Fourteen studies (14\%) focused on three Preparedness, Response, and Recovery. Two studies (1\%) addressed the Preparedness phase. One study (1\%) concentrated on the Mitigation, Preparedness, Response, and Recovery phases, and another (1\%) focused on Preparedness and Response. needless to say, three studies out of 136 (3\%) do not fall into the specified categories of emergency cases.

\subsubsection{Types of UAVs Utilized}

A specific kind of UAV that healthcare professionals utilize for carrying medical supplies is called a Medicare UAV \cite{R043}. We found a limited number of articles mentioning the type of drone they were using. According to the statistical analysis, UAV technology is expected to be valued at over \$600 million in the next few decades. Any other kind of civilian UAV fitted with cameras, sensors, lighting, and other boosted features to facilitate precise navigation, concise data collecting, and minimal communication needs can be considered a Medicare UAV \cite{L104}.

Table \ref{tab:UAVmodelsused}  lists various UAV types used in selected research papers, encompassing a diverse range of unmanned aerial vehicles. Among the featured models are Bebop Drones, Clogworks Dark Matter HX, DJI M600, DJI Mavic 2, DJI Phantom 2, DJI Phantom 4 Pro, DJI S900, DJI M600 Pro, Flirtey's Eagle, and the M1000. This highlights the variety of UAV technologies explored within the literature. Moreover, the utilized UAVs vary not only in terms of design (number of rotors) but also in their intended purposes and functionalities. The inclusion of well-known models such as the DJI Mavic 2, renowned for its compact design and advanced imaging capabilities, along with innovative designs like Flirtey’s Eagle, highlights the wide range of applications of UAVs in different domains.

\begin{table}[t]
\centering
\caption{UAV Types Used in the Analyzed Articles}
\label{tab:UAVmodelsused}
\renewcommand{\arraystretch}{1.5} 
\resizebox{0.90\linewidth}{!}{%
\begin{tabular}{|p{6cm}|p{6cm}|p{6cm}|}
\hline
\textbf{UAV Model}&  \textbf{Use case}& \textbf{Reference}\\
\hline
\hline

Parrot AR-Drone 2.0 & Disaster Monitoring & \cite{L126} \\
\hline
Clogworks Dark Matter HX  & Drug Delivery &  \\
\hline
DJI M600 & Blood Delivery & \cite{L124}\\
\hline
DJI Mavic 2 & Search and Rescue & \cite{L031}\\
\hline
DJI Phantom 2& Lab Sample Delivery & \cite{L128}\\
\hline
DJI Phantom 4 Pro & Damage Assessment and Mapping, Search and Rescue& \cite{L089},\cite{L104}\\
\hline
DJI S900 & Search and Rescue & \cite{L043} \\
\hline
DJIM600 Pro &  Search and Rescue & \cite{L032}\\
\hline
Flirtey's Eagle & AED Delivery& \cite{L020}\\
\hline
M1000 & AED Delivery& \cite{L108}\\
\hline
UX5 & Telemedicine& \cite{L056}\\
\hline
Zipline & Blood& \cite{L028}, \cite{L122}, \cite{L125} \\
\hline
\end{tabular}
}
\end{table}

Identifying the appropriate UAV for healthcare missions requires consideration of several parameters. One of the crucial factors is payload capacity, which directly impacts the UAV's ability to transport medical supplies such as vaccines, blood samples, or essential medications \cite{L124}. The range and endurance of the UAV play a crucial role in determining the distance it can cover and the duration it can remain operational during healthcare missions. Moreover, the UAV's flight capabilities, including stability, maneuverability, and adaptability to different areas and weather conditions, are crucial factors in ensuring reliable and safe delivery of medical supplies \cite{ L128}.

\subsubsection{Advantages of Using UAVs in Healthcare Service}
The integration of UAVs into healthcare and emergency services presents significant advantages, primarily improving response times, increasing survival rates, and ensuring cost-effectiveness. Rapid response times are crucial as they enhance the chances of survival and positive outcomes during emergencies. UAVs contribute to maximizing the benefits of this golden time for medical intervention by quickly delivering essential medical supplies and equipment, thus facilitating prompt and effective treatment \cite{L046, L048}. Studies have highlighted that reducing response times leads directly to increased survival rates, as timely interventions during emergencies are critical \cite{L050, L051}. Additionally, UAVs are recognized for their cost-effectiveness compared to traditional emergency response services like helicopters and ground vehicles. The lower operational costs and reduced need for human intervention contribute to overall financial savings in healthcare logistics \cite{L009, L052}. UAVs can travel long distances in shorter time frames, bypassing traffic and other obstacles, thereby decreasing transportation duration and optimizing the delivery of medical supplies and equipment \cite{L049}.

Moreover, UAVs enhance communication and coordination among emergency responders, patients, and healthcare providers during crises, resulting in quicker and more effective emergency responses  \cite{L050, L051}. The COVID-19 pandemic has underscored the importance of telemedicine, and UAVs can significantly support these efforts by delivering drugs and medical supplies to remote locations. They also facilitate the timely transportation of organs for transplantation and the rapid delivery of AEDs for out-of-hospital cardiac arrest patients, ensuring these critical items reach their destinations swiftly and efficiently \cite{L054}. By minimizing manual tasks, simplifying logistics, and optimizing workflows, UAVs significantly enhance the efficiency and productivity of healthcare operations \cite{L052}.

\section{DISCUSSION}
In this section, we discuss the main findings identified during our analysis of the state of the art, in perspective with the research questions presented earlier in Section \ref{sec:assessmentOfPrimaryStudies}. We then highlight gaps in the state of the art and identify open challenges for future research on using UAVs in healthcare.

\subsection{Research Question Discussion}

\textbf{RQ-1: What are the application domains of UAVs in healthcare services?} We found that UAVs are utilized across nine application domains. Among the eight primary application areas, five are specifically related to healthcare: \textbf{healthcare emergencies}, \textbf{medical supply delivery}, \textbf{telemedicine}, \textbf{telesurgery}, and \textbf{medical surveillance}. 

In healthcare emergencies, UAVs offer rapid response capabilities that significantly enhance the effectiveness of emergency medical services. For instance, the delivery of Automated External Defibrillators (AEDs) for out-of-hospital cardiac arrest (OHCA) is urgent, as immediate access to AEDs can enormously improve survival rates \cite{L113, L114}. In addition, UAVs can transport essential medications, including lifesaving drugs such as Naloxone for opioid overdoses, ensuring that patients receive timely treatment \cite{L115, L116}. Delivering vaccines rapidly is crucial during outbreaks, providing immediate immunization to at-risk populations and controlling the spread of diseases \cite{L065}. Furthermore, UAVs can quickly carry blood samples to laboratories for urgent testing, facilitating prompt diagnosis and treatment decisions \cite{L161}.

In medical supply delivery, UAVs can be used for transporting glucose test kits, pulse oximeters, and automatic sphygmomanometers to remote areas. In cases of severe lower limb bleeding, UAVs can deliver specialized medical supplies, such as tourniquets and hemostatic agents, which are essential for stopping bleeding and stabilizing patients before they reach a medical facility \cite{L140}. 

Telemedicine employs UAVs to support remote medical consultations and follow-ups, significantly enhancing healthcare delivery in remote areas. These UAVs establish communication links that facilitate real-time video consultations, ensuring that patients receive timely medical advice and monitoring without the need for travel.  Furthermore, UAVs play a vital role in the delivery of samples for testing and diagnosis. In remote areas, collecting and transporting medical samples to laboratories can be time-consuming and logistically challenging. UAVs can rapidly transport blood samples, swabs, and other diagnostic materials to centralized laboratories, ensuring that tests are conducted quickly and accurately. This rapid transport reduces the turnaround time for diagnosis, enabling quicker medical interventions and improving patient outcomes. For instance, in outbreak situations, timely diagnosis is critical for containing the spread of infectious diseases. UAVs can accelerate the delivery of diagnostic samples from affected areas to testing facilities, allowing for rapid identification and management of disease outbreaks \cite{L046}, \cite{L056}, \cite{L136}. 

Telesurgery involves using UAVs to perform surgical procedures remotely, an application that offers great benefits, particularly in underserved and geographically isolated regions. This application is especially beneficial in regions lacking specialized surgeons and advanced medical facilities, where timely access to expert surgical care can be life-saving. By utilizing UAVs equipped with precise robotic instruments and advanced communication technologies, expert surgeons can conduct complex operations from afar. These UAVs are outfitted with high-definition cameras, robotic arms, and real-time data transmission capabilities, allowing surgeons to perform delicate procedures with precision and control. In conflict zones or disaster areas, where medical infrastructure is often compromised, telesurgery can play a crucial role. UAVs can rapidly reach these hazardous locations, providing a stable platform for remote surgical interventions. This capability not only bridges the gap between patients and advanced medical care but also minimizes the risks associated with transporting critically ill patients over long distances. Moreover, telesurgery via UAVs ensures that patients receive the necessary surgical care without delay, significantly improving outcomes in emergency situations. The deployment of telesurgery UAVs in such regions also facilitates continuous medical education and training. Local healthcare providers can observe and learn from expert surgeons, enhancing their skills and knowledge. This collaborative approach helps build local medical capacity, ensuring sustainable healthcare improvements in the long term. Additionally, the data collected from these remote surgeries can be used to refine and advance surgical techniques, contributing to the overall progress of medical science \cite{L171}. 

Medical surveillance is another significant domain where UAVs play a crucial role. They are used to monitor the spread of diseases, vaccination drives, and general public health trends. By providing aerial surveillance, UAVs can cover more ground efficiently and offer valuable insights for public health officials \cite{L049}. Humanitarian activities greatly benefit from UAVs, particularly in delivering aid to conflict zones or areas affected by natural disasters. UAVs can bypass ground obstacles, ensuring that medical supplies, food, and other essentials reach affected populations rapidly \cite{L077}. \\

\textbf{RQ-2: What are the main objectives of utilizing UAVs in healthcare?} We analyzed all of the articles and found several objectives, including i) \textbf{improving response time}, ii) \textbf{increasing survival rate}, and iii) \textbf{decreasing cost}.

First, the most crucial objective highlighted across the studies was improving response time. UAVs were recognized as a crucial tool to speed up responses during emergencies, facilitating the delivery of medical supplies and aid to hard-to-reach areas. Compared to urban and rural settings, UAVs arrived before the existing emergency management system in a significant percentage of cases in both urban and rural areas, with considerable time saved, particularly in rural locations. This suggests that UAVs have the potential to significantly reduce response times, especially in areas with extended existing emergency management system response times \cite{L001}. 

Additionally, improving response times is important in out-of-hospital cardio arrest (OHCA) cases, as studies have shown that the chance of survival increases when CPR is provided and first defibrillation is given within the first eight minutes. Many countries have targets for emergency management systems to reach patients within a certain timeframe, but many fail to meet their response-time targets \cite{L013}. Using UAVs to deliver automated external defibrillators (AEDs) can help improve response times for out-of-hospital cardiac arrest (OHCA) situations. It has been suggested that placing AEDs on county land and, in particular, post offices can greatly enhance AED delivery response times, reaching a more significant portion of the population more quickly \cite{L014}. Some studies have highlighted that no single standard method is universally best for all areas. It acknowledges that each area will vary in terms of available land, placement spots, and geographical features, and therefore, the most beneficial placement method will vary based on the specific characteristics of each area \cite{L012}.

Moreover, the method Gino et al used for finding the result involved the development of a customized UAV prototype named PHOENIX, designed to remotely deliver AEDs to patients experiencing sudden cardiac arrest. The study also involved the creation of a simulation scenario based on a rural community with a cottage hospital, a licensed drone pilot, and a team of paramedics with an ambulance. The scenario was designed to test the necessary care pertinent to an out-of-hospital cardiac arrest (OHCA) situation, including guidance about safety, victim responsiveness assessment, and the initiation of high-quality cardiopulmonary resuscitation. A mathematical model was used on more than 53,000 OHCAs in eight response regions to find the best places for standing drones (bases) to deliver AEDs, which showed that response times could be cut down. The study found that using UAVs for AED delivery and other time-critical medical supplies could reduce response time by up to 10.5 minutes in rural settings, potentially leading to improved survival rates for sudden cardiac arrest victims. Additionally, a trial of AED delivery by drone to rural communities in Southern Ontario, Canada, showed that the time to apply an AED to a victim of cardiac arrest decreased by 1.8–8.0 minutes when a drone was utilized. These results suggest that the use of drones for AED delivery has the potential to significantly improve response times in emergency situations \cite{L119}. 

One study used a two-stage approach to find the region-specific UAV network that improved the median regional 911 response time by more than or equal to 1 minute, and this process was repeated for 2 and 3 minutes. For each combination of UAV response time improvement goals 1, 2, and 3 minutes faster than the median 911 response time and region, the number of bases and drones required was quantified. Using the out-of-sample testing set of OHCAs, the response time distribution of the optimized drone network was determined. The study found that UAVs not only improve the median time to defibrillator arrival on the scene but also reduce the entire response time distribution \cite{L072}. 

Furthermore, increasing survival rates emerged as a crucial objective linked to UAV utilization. These aerial vehicles were shown to play a vital role in improving survival rates by enabling rapid access to medical supplies, real-time monitoring, and support in search and rescue missions. UAVs can potentially increase survival rates for OHCA by providing more timely access to early defibrillation. One study highlights that survival rates of 50-70\% from OHCA can be achieved with defibrillation within 3-5 minutes of OHCA, and the probability of survival is reduced by 10\% for each minute delay. However, less than 2\% of victims have an AED applied before the arrival of an ambulance \cite{L015}. The main point about the rising survival rates seen in out-of-hospital cardiac arrest (OHCA) in Singapore and Victoria is that the differences in survival rates between the two areas may be due to differences in how emergency medical services (EMS) choose which patients to resuscitate and transport \cite{L027}. Therefore, the factors that influence survival in resuscitated OHCA patients include age, arrest location, and cardiac etiology \cite{L020}.

The potential impact of using UAVs in search and rescue operations to increase survival rates is significant. By utilizing UAVs to locate victims faster and provide quicker intervention, this technology can be crucial in improving patient outcomes in critical situations \cite{L098}. Reducing the time to reach the victim is especially important in scenarios like hypothermia and cardiac arrest, where rapid intervention is vital for preserving brain function and increasing survival chances \cite{L100}. UAVs can also speed up access to crucial medical equipment like automatic external defibrillators (AEDs), potentially improving patient outcomes in out-of-hospital cardiac arrest situations \cite{L119}. The evolving role of drones in medical applications, particularly in mountain rescue scenarios, suggests positive implications for patient health and survival rates. Overall, the use of drones in search and rescue operations can contribute to saving lives by enabling faster response times, efficient victim location, and timely medical interventions, ultimately leading to increased survival rates in emergency situations \cite{L105}.

 The analysis of these articles emphasized the assessment of cost-effectiveness in healthcare and emergency management scenarios. One study provides insights into the cost-effectiveness of different UAV network coverage levels (80\%, 90\%, and 100\%) in rural areas lacking timely access to emergency management systems. The calculated incremental cost-effectiveness ratio (ICER) for each UAV network indicates that the UAV networks with 80\% and 90\% coverage could be considered cost-effective based on international thresholds\cite{L110}. Some researchers discuss the cost-effectiveness of UAVs in medical services in several contexts. They mentioned that UAV transportation of blood products was found to be more cost-effective than ambulance transportation, with the significantly shorter transport time of the UAV offsetting its cost per minute. Also, some studies highlight the potential for cost savings in smaller communities through the successful uptake of UAV-delivered AEDs \cite{L115}. 
 
Some researchers used simulation modeling to assess the impact of using UAVs for routine vaccine distribution. They used the HERMES (Highly Extensible Resource for Modeling Event-Driven Supply Chains) software platform to develop a discrete-event simulation model of the World Health Organization's (WHO) Expanded Program on Immunization (EPI) supply chain in Gaza, a province in southern Mozambique. The results showed that implementing a UAV could increase vaccine availability and decrease costs in a wide range of settings and circumstances if the UAVs are used frequently enough to overcome the capital costs of installing and maintaining the system. The UAV maintained cost savings in all sensitivity analyses, ranging from 0.05 to 0.21 per dose administered. This study found that the UAV offered cost savings through lower transport, per diem, and labor costs that offset the additional hub infrastructure costs \cite{L074}.

Some studies compared the costs of UAVs versus ambulances for delivering blood products to treat maternal obstetric hemorrhages in challenging terrain and traffic conditions\cite{L076}. The economic evaluation concluded that although UAV transportation of blood products costs more compared to ambulance transport, the significantly reduced travel time offsets the cost \cite{L074}. From an economic viewpoint, UAVs are considered a more cost-effective and viable mode of blood product transportation, particularly during emergencies. These findings contribute to the understanding of the cost-effectiveness of drones in healthcare service delivery, where delivery time is crucial.\cite{L150}.

\textbf{RQ-3: What are the main challenges in adopting UAVs for enhancing healthcare services?} The main challenges considered in most works are: i) ensuring the safety of both the UAVs and the people they have connected with; ii) weather conditions; iii) laws and regulations; iv) battery life; and finally, v) payload temperature and stability.

Ensuring the safety of UAVs and the people they interact with is a primary concern in their integration into healthcare services. This requires developing reliable collision-avoidance systems, implementing operational protocols, and ensuring UAVs can fly safely in populated areas, such as urban areas, hospitals, and other medical facilities to prevent collisions \cite{L017}, \cite{L043}. Additionally, the human factor in UAV interactions should be carefully considered, emphasizing the need for training and awareness to minimize the risk of accidents. Weather conditions pose another significant challenge, as adverse weather such as high winds, rain, or extreme temperatures can impact UAV performance and compromise the safety of medical payloads. Researchers are exploring advanced weather monitoring systems and predictive algorithms to enhance UAV navigation in challenging conditions \cite{L004, L040, L122}.

Legal and regulatory frameworks are critical to the lawful operation of medical UAVs. This involves ensuring UAVs operate within the framework of national and international regulations, which include considerations for air traffic management and airspace regulations to ensure their safe coexistence with conventional air traffic \cite{L009, L015, L018}. Another major challenge is battery life, as limited battery capacity restricts the operational range and duration of medical UAV missions. Ongoing research focuses on developing advanced battery technologies, such as lightweight and high-capacity batteries, and exploring efficient energy management systems \cite{L007}. Furthermore, UAV operations must consider payload temperature and stability, especially during the delivery of sensitive items like organs or blood samples. Ensuring the temperature and stability of UAVs during such missions is critical to maintaining the integrity of medical deliveries \cite{L150, L051}.

\textbf{RQ-4: What UAV types are used for healthcare and services?} Out of 136 articles, 16 explicitly mentioned the specific UAV models utilized in their studies. The diverse range of UAVs identified in the literature contains various technologies and configurations that include the 8-Rabbit AR-Drones Clogworks Dark Matter HX, DJI M600, DJI Mavic 2, DJI Phantom 2, DJI Phantom 4 Pro, DJI S900, DJI M600 Pro, Flirtey’s Eagle, and the M1000. This list emphasizes the variety of UAV technologies explored within the context of healthcare services. Notably, these UAVs show features and capabilities reflecting the evolving landscape of unmanned aerial systems in healthcare applications.

\subsection{Research Gaps, and an Agenda for Future Work}

The outcomes of this systematic review revealed that UAVs have great potential for different applications in the healthcare sector, as appears from the great increase in the number of publications within the last three years of this study. This notable increase in publications over the past three years also indicates a growth in knowledge and awareness regarding UAV applications in healthcare and emergency scenarios. However, a significant part of the existing literature is based on simulation, which introduces a gap in UAV adoption in real-world use cases. 

Regarding the identified use cases, about $40\%$ of the applications have been employed to focus on healthcare emergencies, and then about $33\%$ of the focus was on disaster management. However, the systematic analysis revealed that some use cases still have a limited number of publications or research papers, including medical supply delivery, medical surveillance, environmental monitoring, telemedicine, and telesurgery. 
 
The successful and timely delivery of organs for transplantation or vaccines to remote locations or hard-to-reach areas is critical for public health \cite{L109}. Further research in this area can lead to innovations in delivery methods, protection techniques, and supply chain management, ultimately improving the accessibility and effectiveness of healthcare services \cite{L110}. Similarly, increasing situational awareness in disaster areas and evacuation strategies are vital for minimizing losses and responding effectively to emergencies. Additionally, there is a lack of work on UAV utilization in the post-disaster period (e.g., victim identification) \cite{L014}.

The existing literature primarily addresses the use of UAVs under normal weather conditions, often neglecting the complexities introduced by real disaster scenarios, which can lead to operational delays \cite{L050}. To address these challenges, researchers should incorporate considerations of adverse weather conditions into their studies, with particular emphasis on waterproofing electronic components to ensure functionality during rainy weather. Furthermore, UAV-specific limitations, such as battery life, become critical when handling heavy payloads and covering long distances. However, decision models for UAV adoption across diverse conditions are currently lacking. 
Decision models are essential for determining when and how UAVs should be used. Some studies have started to propose frameworks for integrating UAVs into the chain of survival, particularly emphasizing the need for standardized procedures to guide healthcare professionals \cite{L136}. Creating comprehensive decision models will be crucial to move beyond pilot cases and ensure the safe, effective, and routine use of UAVs in medical emergencies. This will involve developing guidelines for risk assessment, operational procedures, and integration with existing healthcare infrastructures to fully leverage the potential of UAV technology.

While UAVs can be necessary for search and rescue missions in flooded areas where ground vehicles cannot operate, they might not be as effective for continuous surveillance due to limited battery life. Moreover, studies show UAVs can significantly reduce response times and improve outcomes in rural settings, making them a cost-effective solution compared to traditional ambulance services \cite{L124}. However, in urban areas with well-established emergency services, the benefits may not justify the costs unless specific logistical advantages are present. Thus, the critical evaluation of scenarios where UAVs offer a clear advantage over traditional methods is essential, recognizing that no universal solution exists and that the effectiveness of UAVs depends significantly on the specific context and operational requirements.

Furthermore, low awareness and negative perceptions surrounding the importance of UAVs in healthcare and emergency response within local communities, particularly in rural areas, could be barriers to UAV adoption \cite{L081}. Thus, training and increasing awareness are essential for the successful adoption of UAVs. 

While the literature provided some information on the UAV models, it is notable that a considerable number of articles did not mention the UAV type employed in their studies. This observation highlights an opportunity for further research to bridge the gap between UAV types and their applications in healthcare services. Creating a mapping between UAV types and their specific healthcare applications facilitates more informed decisions in deploying UAVs for healthcare purposes. Furthermore, an essential aspect that emerged from our analysis is the need for additional research on the technical appropriateness of each UAV model in the healthcare context. Investigating the technical specifications, capabilities, and limitations of these UAVs in the healthcare domain is crucial for ensuring their effective and safe integration into healthcare services.

The security aspect of UAVs in healthcare extends beyond their physical capabilities to include medical device protection, medical supply protection, and privacy \cite{L149}. As these aerial vehicles collect and transfer sensitive medical information, ensuring robust cybersecurity measures is crucial to safeguarding medical supplies and devices and protecting the integrity of healthcare data. By prioritizing both the physical limitations and security, integrating these technologies into medical healthcare systems can be facilitated, creating a safer and more efficient healthcare infrastructure for all.

Table \ref{tab:research_agenda} outlines the key research questions that need to be addressed to drive future advancements in the field.

\begin{table}[t]
\centering
\caption{Research Agenda}
\label{tab:research_agenda}

\renewcommand{\arraystretch}{1.5} 
\resizebox{1.00\linewidth}{!}{%
\begin{tabular}{|p{0.65cm}|p{6.5cm}|p{14cm}|}
\hline
\textbf{No} & \textbf{Research Area} & \textbf{Research Question} \\
\hline
\hline
1 & Real-world Implementations & How can UAVs be effectively integrated into real-world healthcare services? And how can UAV technologies be further developed to expand their use in underrepresented healthcare applications, such as medical supply delivery and telemedicine? \\ \hline

4 & Safe and Smooth Integration & How can UAVs be seamlessly integrated with existing healthcare infrastructures to enhance healthcare and emergency services in a safe and effective manner? \\ \hline

2 & Comprehensive Decision Models & What comprehensive decision models can be developed to optimize the integration of UAVs into emergency medical services, ensuring their safe and effective use under various conditions? \\ \hline

3 & Reliability and Efficiency & What innovations in UAV technology are needed to enhance the reliability and efficiency of delivering critical medical supplies to remote and hard-to-reach areas, particularly under adverse weather conditions? \\ \hline

5 & Security Measures & What security measures and regulations are required to protect sensitive medical data when using UAVs in healthcare, and how can these be effectively implemented? \\ \hline

8 & Cost-Effectiveness & In what scenarios do UAVs offer significant advantages over traditional methods in terms of cost and effectiveness? \\ \hline

6 & Community Awareness and Training & How can awareness and perceptions of UAVs in healthcare be improved to enhance their adoption? \\ \hline

\end{tabular}
}
\end{table}

\section{CONCLUSION}
This article presents a systematic review of the utilization of UAVs in healthcare services, analyzing 136 papers over three decades to address four primary questions: (i) the applications of UAVs in healthcare services, (ii) the types of UAVs used, (iii) the advantages of UAVs, and (iv) the challenges within this domain. Our analysis provides a comprehensive understanding of UAV adoption in healthcare, clarifying their typical characteristics and implications. The study determines several advantages of UAVs, such as improving response times, increasing survival rates, and enhancing cost-effectiveness. However, many of these benefits are still at the conceptual or pilot stages, needing further evidence to convince decision-makers. The advantages are often reported from a high-level perspective, requiring more specific use case analysis. For instance, while UAVs can theoretically improve response times, this does not guarantee higher survival rates, as inappropriate UAV deployment could even lead to lower survival rates. This emphasizes the need for robust decision protocols in healthcare settings. Future research should focus on detailed case studies and empirical evidence to reinforce the practical benefits of UAVs and develop standardized protocols for their effective implementation in healthcare services.

Some limitations must also be reported in the systematic review of UAV applications in healthcare. Firstly, the scope of sources used was restricted, indicating that other relevant databases might have been overlooked, potentially leading to an incomplete representation of the literature. Additionally, the novelty of UAV applications in healthcare means that many studies are conceptual rather than empirical, limiting the available data on real-world applications and outcomes. The review also identified clear gaps and research challenges for future work, primarily related to (i) regulatory considerations, which are crucial for the safe and legal integration of UAVs into healthcare systems, (ii) ensuring the security and safety of UAVs in medical contexts to protect patient privacy and prevent uncertified access or malfunctions, (iii) technology acceptance among healthcare providers and patients, which is essential for successful implementation, and (iv) specifying UAV types and their technical characteristics, and establishing a mapping between UAV types and their applications in healthcare to optimize their use and improve efficiency. Future research should expand to include the broader ecosystem involved in UAV healthcare applications, examining the roles and evolution of suppliers, infrastructure, maintenance, and the emerging market. Additionally, unifying studies involving healthcare professionals could provide insights into practical and technical challenges, further enriching the discourse on UAV integration in medical contexts.


\bibliographystyle{IEEEtran}  
\bibliography{references}     

\end{document}